\RequirePackage[l2tabu, orthodox]{nag}
\RequirePackage{snapshot}

\documentclass[11pt,onecolumn]{article}

\usepackage[dvips,letterpaper,top=0.5in, bottom=0.5in, left=0.75in, right=0.5in,includefoot,heightrounded]{geometry}

\usepackage{graphicx}
\usepackage{amsmath}
\usepackage{amssymb,amsfonts}

\usepackage{abstract}

\usepackage[usenames,dvipsnames]{color}
\usepackage[usenames,dvipsnames]{xcolor}
\usepackage{subfigure}

\usepackage{booktabs}

\usepackage{setspace}
\usepackage{flushend}
\usepackage{multicol}

\usepackage{cite}
\usepackage{url}
\usepackage[normalem]{ulem}

\usepackage{enumerate}

\usepackage{psfrag}

\usepackage{bm}

\usepackage[sc,tiny]{titlesec}
       \usepackage{mathptmx}



\title{%
A Class of DCT Approximations Based on the Feig-Winograd Algorithm
}

\author{%
C.~J.~Tablada%
\thanks{%
C. J. Tablada and R.~J.~Cintra are with 
the Signal Processing Group,
Departamento de Estat\'{\i}stica, 
Universidade Federal de Pernambuco.
E-mail: 
\protect\url{rjdsc@dsp.ufpe.org}
}
\and
F. M. Bayer%
\thanks{%
F. M. Bayer
is with the
Departamento de Estat\'{\i}stica and LACESM, 
Universidade Federal de Santa Maria, Brazil.
}
\and
R.~J.~Cintra${}^\ast$%
}

\date{}

\newcommand{\myabstract}{%
A new class of
matrices based on a parametrization of
the Feig-Winograd 
factorization of $8$-point DCT is proposed.
Such 
parametrization induces a matrix subspace,
which unifies a number of existing methods
for DCT approximation.
By solving a comprehensive
multicriteria optimization problem,
we identified several new DCT approximations.
Obtained solutions were sought
to possess the following properties:
(i)~low multiplierless
computational complexity,
(ii)~orthogonality or near orthogonality,
(iii)~low complexity invertibility,
and
(iv)~close proximity and performance to the exact DCT.
Proposed approximations
were submitted to assessment in terms
of 
proximity to the DCT,
coding performance,
and
suitability for image compression.
Considering
Pareto efficiency,
particular new proposed approximations
could outperform various existing methods archived in literature.
}

\newcommand{\mykeywords}{%
DCT approximation,
Feig-Winograd algorithm,
Low-complexity transforms,
Image compression
}

\begin{document}

\makeatletter
\if@twocolumn

\twocolumn[%
  \maketitle
  \begin{onecolabstract}
    \myabstract
  \end{onecolabstract}
  \begin{center}
    \small
    \textbf{Keywords}
    \bigskip
    \linebreak

    \mykeywords
  \end{center}
  \bigskip
]
\saythanks

\else

  \maketitle
  \begin{abstract}
    \myabstract
  \end{abstract}
  \begin{center}
    \small
    \textbf{Keywords}
    \bigskip
    \linebreak
    \mykeywords
  \end{center}
  \bigskip
  \onehalfspacing
\fi

\section{Introduction}

The discrete cosine transform (DCT) 
is an essential tool in 
digital signal processing~\cite{rao1990discrete, britanak2007discrete}.
This is mainly because
the DCT is asymptotically equivalent to
the
Karhunen-Lo\`eve transform~(KLT),
which
possesses
optimal decorrelation and energy compaction
properties~\cite{ahmed1974, Clarke1981, rao1990discrete, britanak2007discrete, Liang2001, haweel2001new}.
Such good characteristics are attained 
when
high correlated
first-order Markov signals are considered~\cite{rao1990discrete, britanak2007discrete}.
Importantly
natural images belong to 
this particular class of signals~\cite{Liang2001}.

In particular,
the
DCT
has been adopted in several
image and video coding 
schemes~\cite{bhaskaran1997},
such as 
JPEG~\cite{Wallace1992},
MPEG-1~\cite{roma2007hybrid}, 
MPEG-2~\cite{mpeg2}, 
H.261~\cite{h261}, 
H.263~\cite{h263}, 
ACV/H.264~\cite{wiegand2003}, %
and 
the recent HEVC/H.265~\cite{hevc1}.
In particular, 
H.264 and HEVC standards employ
8-point DCT algorithms~\cite{Richardson2011, %
Lee2008, %
Mehrabi2010,He2008,Moon2010,
hevc1,sullivan2012,%
Edirisuriya2012} 
among other different blocklenghts, such as 4, 16, and 32~points~\cite{Park2012, Park2013}.
In~\cite{Potluri2013},
the 8-point DCT stage of the HEVC 
was optimized.
Because of this increasing demand,
several algorithms for the efficient computation
of 
the 8-point DCT
have been proposed~\cite{vetterli1984simple, hou1987fast}.
In~\cite{rao1990discrete,britanak2007discrete},
comprehensive surveys on DCT algorithms are detailed.

Noteworthy DCT methods
include
the following procedures:
Wang factorization~\cite{wang1984fast},
Lee DCT for power-of-two block lengths~\cite{lee1984new},
Arai DCT scheme~\cite{arai1988fast},
Loeffler algorithm~\cite{loeffler1991practical},
and
Feig-Winograd factorization~\cite{fw1992}.
All these algorithms are classical results in the field
and
have been considered 
for practical applications~\cite{roma2007hybrid, Vasudev1998, Lin2006}.
For instance,
the Arai DCT scheme was employed in various recent
hardware implementations of the DCT~\cite{arjunamadanayake2011algebraic, rajapaksha2013asynchronous, edirisuriya2012vlsi}.

Indeed,
after
the introduction of the DCT
by Ahmed~\emph{et al.}~\cite{ahmed1974}
in 1974,
designing efficient DCT algorithms has been
a major scientific effort
in the circuits, systems, and signal processing community.
Because of such
intense research in the field,
the current \emph{exact} 
methods are very close to 
the theoretical DCT 
complexity~\cite{Liang2001,
Madanayake2013, 
Heideman1988,
loeffler1991practical,
arai1988fast}.
Therefore,
it may be unrealistic 
to expect that
new exact
algorithms could offer dramatic computational gains
for such a fundamental and deeply investigated mathematical operation.

In this scenario,
approximate methods 
offer
an alternative way to further
reduce the computational complexity
of the DCT~\cite{britanak2007discrete,
haweel2001new,
Chen2002,
bc2012,
bas2008,
bas2011}.
While not computing the DCT
exactly, such approximations can provide 
meaningful estimations at very low-complexity computational 
requirements.
In this sense, 
literature
has been populated 
with approximate methods for the efficient computation 
of the 
DCT. 
For example, 
the AVC/H.264
and HEVC/H.265 standards employ integer approximate DCT
in order to reduce the computational cost of the transform stage.
A comprehensive list of approximate methods for the DCT 
is found in~\cite{britanak2007discrete}.
Prominent methods include
the
signed DCT (SDCT)~\cite{haweel2001new},
the
level 1 approximation by 
Lengwehasatit-Ortega~\cite{lengwehasatit2004scalable},
the
Bouguezel-Ahmad-Swamy (BAS) series of algorithms~\cite{bas2008, Bouguezel2008, bas2009, bas2010, bas2011, bas2013},
the
DCT round-off approximation~\cite{cb2011},
the 
modified DCT round-off approximation~\cite{bc2012},
and
the
multiplier-free DCT approximation for 
radio-frequency (RF)
multi-beam digital aperture-array space
imaging~\cite{multibeam2012}.

Although several other types of approximations
are available,
in general, 
\emph{very low-complexity}
approximation 
matrices
have
their
elements
defined on
the set 
$\{ 0, \pm 1/2, \pm1, \pm2\}$~\cite{bc2012, cb2011, haweel2001new, bas2011, bas2013}.
Thus,
such
transformations 
possess null multiplicative complexity,
because the required arithmetic operations
can be implemented 
exclusively by means of
binary additions and bit-shifting operations.
Indeed,
DCT approximations 
can replace the \emph{exact} DCT 
in 
hardware implementation and 
high-speed computation/processing~\cite{britanak2007discrete},
while
having
low hardware costs and low power demands~\cite{edirisuriya2012vlsi}.
Effectively,
DCT approximations
have been considered
for applications in
real-time video transmission and processing~\cite{Kuo2011, Saponara2012},
satellite communication systems~\cite{britanak2007discrete},
portable computing applications~\cite{britanak2007discrete},
radio-frequency smart antenna array~\cite{multibeam2012},
and
wireless image sensor networks~\cite{Lecuire2012}.

The proposed methods archived in literature
for generating \emph{very} low complexity 
approximations
include:
(i)~crude approximations~\cite{haweel2001new,cb2011,bas2013};
(ii)~inspection~\cite{bas2008,bas2009,bas2010};
(iii)~variations of previous approximations via a single-parameter matrix~\cite{bas2011};
and
(iv)~optimization procedures based on the DCT structure~\cite{multibeam2012}.
Thus,
the existing approximations appear
as \emph{isolate} cases
without a unifying mathematical formalism.

The aim of this paper is two-fold.
Our first 
goal
is 
to
introduce a new 
class of DCT approximations.
For such end, 
we consider a parametrization of the Feig-Winograd factorization of
the 8-point DCT matrix~\cite{fw1992}.
Such parametrization allows the constructions of 
a specific matrix subspace.
Second, 
over the introduced subspace,
we solve a constrained multicriteria optimization problem
to identifying optimal DCT approximations
according to several figures of merit
for image compression.
Both orthogonal and non-orthogonal approximations
are sought.

The paper unfolds as follows. 
In Section~\ref{section-feig-winograd},
we describe the mathematical structure of
the
proposed class of transforms,
including fast algorithms
for 
the direct and inverse transformations.
Section~\ref{section-approximation} 
discusses
the 
desirable properties
that approximate $8$-point DCTs 
are expected to satisfy,
such as
low computational complexity,
orthogonality,
invertibility,
and
proximity to the exact DCT.
In Section~\ref{S:results},
we
formalize
a multicriteria optimization problem
based on a comprehensive set of performance measures
in order to identify 
efficient solutions and new transformations 
over the proposed
class of matrices.
The resulting
approximations
are subject to assessment and 
extensive comparison
with 
competing methods.
In Section~\ref{S:image-compression}, 
we perform a comprehensive image compression 
analysis considering the 
obtained 
optimal transforms.
using image quality measures as figures of merit.
Section~\ref{S:conclusion}
concludes the paper.

\section{Feig-Winograd Approximate DCT}
\label{section-feig-winograd}

\subsection{Preliminaries}

The DCT is algebraically represented by
the $N \times N$ transformation matrix 
$\mathbf{C}_N$
whose elements are given by~\cite{rao1990discrete, britanak2007discrete}:
\begin{align*}
c_{m,n}= 
\sqrt{\frac{2}{N}}
\,
\beta_{m-1}
\cos
\left(
\frac{\pi (m-1)(2n-1)}{2N} 
\right)
, 
\end{align*} 
where $m,n=1,2,\ldots,N$, 
$\beta_0 = 1/\sqrt{2}$,
and
$\beta_k = 1$,
for $k\neq0$.
Let
$\mathbf{x}=\begin{bmatrix} x_0 & x_1 & \cdots & x_{N-1} \end{bmatrix}^\top$
be an input vector,
where the superscript ${}^\top$ denotes the transposition operation.
The one-dimensional (\mbox{1-D}) DCT transform of~$\mathbf{x}$
is 
the $N$-point vector~$\mathbf{X}=\begin{bmatrix} X_0 & X_1 & \cdots & X_{N-1} \end{bmatrix}^\top$
given by
$\mathbf{X}=\mathbf{C}_N \cdot \mathbf{x}$.
Because $\mathbf{C}_N$ is an orthogonal matrix,
the inverse transformation can be written according to 
$\mathbf{x}=\mathbf{C}_N^{\top} \cdot \mathbf{X}$.

Let $\mathbf{A}_N$ and $\mathbf{B}_N$ be
square matrices
of
order~$N$.
For two-dimensional (\mbox{2-D}) signals,
we have the following relationships.
The forward and inverse \mbox{2-D} DCT operations
are expressed by
\begin{align}
\mathbf{B}_N 
=
\mathbf{C}_N \cdot \mathbf{A}_N \cdot \mathbf{C}_N^{\top}
\quad
\text{and}
\quad
\mathbf{A}_N 
=
\mathbf{C}_N ^{\top} \cdot \mathbf{B}_N \cdot \mathbf{C}_N
, 
\nonumber
\end{align}
respectively.

Although the procedures described in this work
can be applied to any blocklength,
we focus exclusively on the 8-point DCT
which is denoted as~$\mathbf{C}_8$ and is given by:
\begin{align*}
&\mathbf{C}_8  = 
\frac{1}{2}
\cdot
\begin{bmatrix}
\begin{smallmatrix}
\gamma_3 & \phantom{-}\gamma_3 & \phantom{-}\gamma_3 & \phantom{-}\gamma_3 & \phantom{-}\gamma_3 & \phantom{-}\gamma_3 & \phantom{-}\gamma_3 & \phantom{-}\gamma_3 \\
\gamma_0 & \phantom{-}\gamma_2 & \phantom{-}\gamma_4 & \phantom{-}\gamma_6 & -\gamma_6 & -\gamma_4 & -\gamma_2 & -\gamma_0 \\
\gamma_1 & \phantom{-}\gamma_5 & -\gamma_5 & -\gamma_1 & -\gamma_1 & -\gamma_5 & \phantom{-}\gamma_5 & \phantom{-}\gamma_1 \\
\gamma_2 & -\gamma_6 & -\gamma_0 & -\gamma_4 & \phantom{-}\gamma_4 & \phantom{-}\gamma_0 & \phantom{-}\gamma_6 & -\gamma_2 \\
\gamma_3 & -\gamma_3 & -\gamma_3 & \phantom{-}\gamma_3 & \phantom{-}\gamma_3 & -\gamma_3 & -\gamma_3 & \phantom{-}\gamma_3 \\
\gamma_4 & -\gamma_0 & \phantom{-}\gamma_6 & \phantom{-}\gamma_2 & -\gamma_2 & -\gamma_6 & \phantom{-}\gamma_0 & -\gamma_4 \\
\gamma_5 & -\gamma_1 & \phantom{-}\gamma_1 & -\gamma_5 & -\gamma_5 & \phantom{-}\gamma_1 & -\gamma_1 & \phantom{-}\gamma_5 \\
\gamma_6 & -\gamma_4 & \phantom{-}\gamma_2 & -\gamma_0 & \phantom{-}\gamma_0 & -\gamma_2 & \phantom{-}\gamma_4 & -\gamma_6
\end{smallmatrix}
\end{bmatrix}
,
\end{align*}
where
$\gamma_k = \cos(2\pi (k+1) /32)$.

\subsection{Feig-Winograd DCT Factorization}

In~\cite{fw1992}
Feig and Winograd
introduced a fast algorithm for 
the \mbox{1-D} 8-point DCT,
whose factorization can be given by:
\begin{align*}
\mathbf{C}_8
=\frac{1}{2}
\cdot
\mathbf{P}_8
\cdot
\mathbf{K}_8 
\cdot 
\mathbf{B}^{(1)}_8 \cdot \mathbf{B}^{(2)}_8 \cdot \mathbf{B}^{(3)}_8,
\end{align*}
where
$\mathbf{P}_8$ is a signed permutation matrix,
$\mathbf{K}_8$ is a multiplicative matrix,
and
$\mathbf{B}^{(1)}_8$,
$\mathbf{B}^{(2)}_8$,
and
$\mathbf{B}^{(3)}_8$
are 
symmetric 
additive matrices.
These matrices are given by:
\begin{align*}
\mathbf{B}^{(1)}_8 =
\operatorname{bdiag}
\left(
\left[\begin{smallmatrix}1&\phantom{-}1\\1&-1\end{smallmatrix}\right]
,
\mathbf{I}_6
\right)
,\quad
\mathbf{B}^{(2)}_8 =
\operatorname{bdiag}
\left(
\left[\begin{smallmatrix}1&\phantom{-}0&\phantom{-}0&\phantom{-}1\\0&\phantom{-}1&\phantom{-}1&\phantom{-}0\\0&\phantom{-}1&-1&\phantom{-}0\\1&\phantom{-}0&\phantom{-}0&-1\end{smallmatrix}\right]
,
\mathbf{I}_4
\right)
,
\end{align*}
\begin{align*}
\mathbf{B}^{(3)}_8 =
\begin{bmatrix}
\mathbf{I}_4 & \phantom{-}\bar{\mathbf{I}}_4 \\
\bar{\mathbf{I}}_4 & -\mathbf{I}_4
\end{bmatrix},
\quad
\mathbf{P}_8 &=
\begin{bmatrix}
\begin{smallmatrix}
1 &\phantom{-}0 &\phantom{-}0 & \phantom{-}0 & \phantom{-}0 &\phantom{-}0 &\phantom{-}0 &\phantom{-}0 \\
0 &\phantom{-}0 &\phantom{-}0 & \phantom{-}0 & -1 &\phantom{-}0 &\phantom{-}0 &\phantom{-}0 \\
0 &\phantom{-}0 &\phantom{-}1 & \phantom{-}0 & \phantom{-}0 &\phantom{-}0 &\phantom{-}0 &\phantom{-}0 \\
0 &\phantom{-}0 &\phantom{-}0 & \phantom{-}0 & \phantom{-}0 &-1 &\phantom{-}0 &\phantom{-}0 \\
0 &\phantom{-}1 &\phantom{-}0 & \phantom{-}0 & \phantom{-}0 &\phantom{-}0 &\phantom{-}0 &\phantom{-}0 \\
0 &\phantom{-}0 &\phantom{-}0 & \phantom{-}0 & \phantom{-}0 &\phantom{-}0 &\phantom{-}0 &-1 \\
0 &\phantom{-}0 &\phantom{-}0 & \phantom{-}1 & \phantom{-}0 &\phantom{-}0 &\phantom{-}0 &\phantom{-}0 \\
0 &\phantom{-}0 &\phantom{-}0 & \phantom{-}0 & \phantom{-}0 &\phantom{-}0 &\phantom{-}1 &\phantom{-}0
\end{smallmatrix}
\end{bmatrix}
,
\end{align*}
and
\begin{align*}
\mathbf{K}_8 &=
\begin{bmatrix}
\begin{smallmatrix}
\gamma_3 &\phantom{-}0 &\phantom{-}0 & \phantom{-}0 & \phantom{-}0 &\phantom{-}0 &\phantom{-}0 &\phantom{-}0 \\
0 & \gamma_3 &\phantom{-}0 & \phantom{-}0 & \phantom{-}0 &\phantom{-}0 &\phantom{-}0 &\phantom{-}0 \\
0 &\phantom{-}0 & \phantom{-}\gamma_5 & \phantom{-}\gamma_1 & \phantom{-}0 &\phantom{-}0 &\phantom{-}0 &\phantom{-}0 \\
0 &\phantom{-}0 & -\gamma_1 & \phantom{-}\gamma_5 & \phantom{-}0 &\phantom{-}0 &\phantom{-}0 &\phantom{-}0 \\
0 &\phantom{-}0 &\phantom{-}0 & \phantom{-}0 & -\gamma_6 & -\gamma_4 & -\gamma_2 & -\gamma_0 \\
0 &\phantom{-}0 &\phantom{-}0 & \phantom{-}0 & \phantom{-}\gamma_4 & \phantom{-}\gamma_0 & \phantom{-}\gamma_6 & -\gamma_2 \\
0 &\phantom{-}0 &\phantom{-}0 & \phantom{-}0 & -\gamma_0 & \phantom{-}\gamma_2 & -\gamma_4 & \phantom{-}\gamma_6 \\
0 &\phantom{-}0 &\phantom{-}0 & \phantom{-}0 & -\gamma_2 & -\gamma_6 & \phantom{-}\gamma_0 & -\gamma_4 \\
\end{smallmatrix}
\end{bmatrix}
,
\end{align*}
where 
$\mathbf{I}_l$ and $\bar{\mathbf{I}}_l$
denote the identity and counter-identity 
matrices of order~$l$,
respectively,
and $\operatorname{bdiag}(\cdot)$ 
is the block diagonal operator.

Above factorization
circumscribes
the entire multiplicative complexity
of the DCT 
into the block diagonal matrix~$\mathbf{K}_8$.
Indeed,
the seven distinct non-null elements of~$\mathbf{K}_8$,
namely $\gamma_i$, $i=0,1,\ldots,6$,
are the only non-trivial quantities
in Feig-Winograd algorithm.

\subsection{Feig-Winograd Matrix Mapping}

The Feig-Winograd factorization
paves the way for
defining a class of 8$\times$8 matrices 
generated according to the following
mapping:
\begin{align}
\operatorname{FW}:
\mathbb{R}^7
&
\longrightarrow \mathcal{M}(8) 
\nonumber
\\
\bm{\alpha} 
&
\longmapsto 
\mathbf{P}_8
\cdot
\mathbf{K}^{(\bm{\alpha})}_8
\cdot 
\mathbf{B}^{(1)}_8 \cdot \mathbf{B}^{(2)}_8 \cdot \mathbf{B}^{(3)}_8,
\label{e:mat_fast}
\end{align}
where
$\bm{\alpha} = 
\begin{bmatrix}\alpha_0 & \alpha_1 & \cdots & \alpha_6 \end{bmatrix}^\top$ 
is a 7-point parameter vector,
$\mathcal{M}(8)$ is the 8$\times$8 matrix space over the real numbers,
and
\begin{align}
\label{e:K_alpha}
\mathbf{K}^{(\bm{\alpha})}_8
&
=
\begin{bmatrix}
\begin{smallmatrix}
\alpha_3 &\phantom{-}0 &\phantom{-}0 & \phantom{-}0 & \phantom{-}0 &\phantom{-}0 &\phantom{-}0 &\phantom{-}0 \\
0 & \alpha_3 &\phantom{-}0 & \phantom{-}0 & \phantom{-}0 &\phantom{-}0 &\phantom{-}0 &\phantom{-}0 \\
0 &\phantom{-}0 & \phantom{-}\alpha_5 & \phantom{-}\alpha_1 & \phantom{-}0 &\phantom{-}0 &\phantom{-}0 &\phantom{-}0 \\
0 &\phantom{-}0 & -\alpha_1 & \phantom{-}\alpha_5 & \phantom{-}0 &\phantom{-}0 &\phantom{-}0 &\phantom{-}0 \\
0 &\phantom{-}0 &\phantom{-}0 & \phantom{-}0 & -\alpha_6 & -\alpha_4 & -\alpha_2 & -\alpha_0 \\
0 &\phantom{-}0 &\phantom{-}0 & \phantom{-}0 & \phantom{-}\alpha_4 & \phantom{-}\alpha_0 & \phantom{-}\alpha_6 & -\alpha_2 \\
0 &\phantom{-}0 &\phantom{-}0 & \phantom{-}0 & -\alpha_0 & \phantom{-}\alpha_2 & -\alpha_4 & \phantom{-}\alpha_6 \\
0 &\phantom{-}0 &\phantom{-}0 & \phantom{-}0 & -\alpha_2 & -\alpha_6 & \phantom{-}\alpha_0 & -\alpha_4 \\
\end{smallmatrix}
\end{bmatrix}
.
\end{align}

The image of the multivariate function $\operatorname{FW}(\cdot)$
is a subset of $\mathcal{M}(8)$.
It is straightforward to verify
that such subset
is also closed under the operations of addition and scalar
multiplication.
Therefore,
this subset is a matrix subspace,
which we refer to as 
the Feig-Winograd matrix subspace.

Mathematically,
the Feig-Winograd factorization induces a 
matrix subspace 
by allowing its multiplicative constants
to be treated as parameters. 
Thus,
an
appropriate parameter selection
may result in suitable approximations.
Our ultimate goal is to identify
in this subspace
matrices
that could adequately
approximate the DCT matrix.
Hereafter
we adopt the following notation
$\mathbf{T}^{(\bm{\alpha})}_8=\operatorname{FW}(\bm{\alpha})$.

\begin{figure}
\centering
\subfigure[Full diagram of the $\mathbf{T}^{(\bm{\alpha})}_8$]{\includegraphics{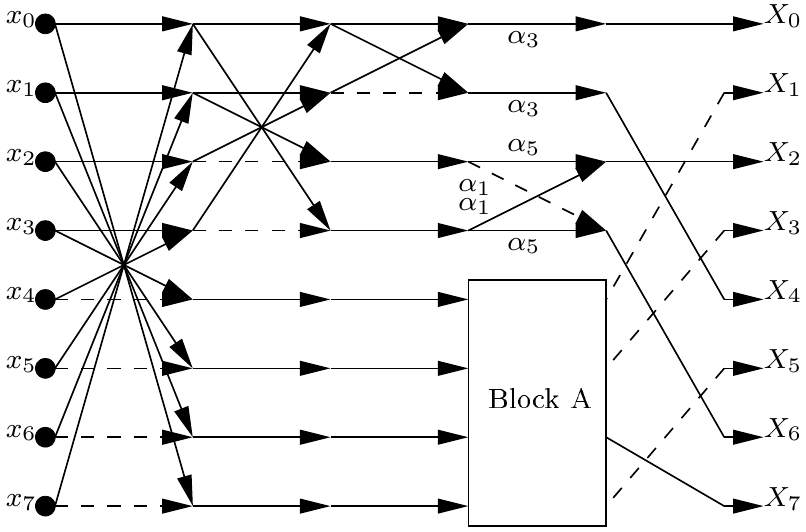}}
\\
\subfigure[Block A]{\includegraphics{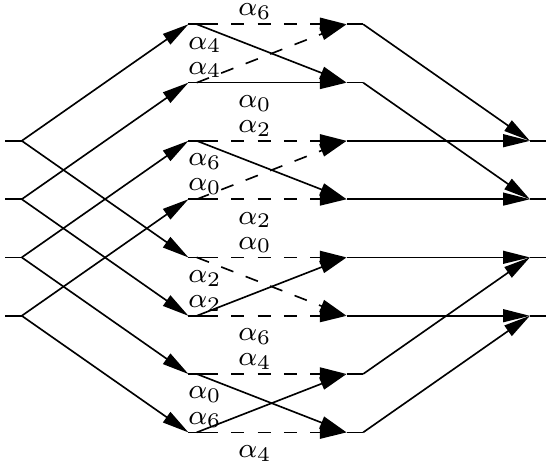}}
\caption{
Signal flow graph for 
the transformations defined on the
Feig-Winograd matrix space.
Input data $x_n$, $n=0,1,\ldots,7$, 
relates to output $X_m$, $m=0,1,\ldots,7$, 
according to $\mathbf{X} = \mathbf{T}^{(\bm{\alpha})}_8 \cdot \mathbf{x}$.
Dashed arrows represent multiplications by $-1$.}
\label{f:fw_fast}
\end{figure}

Considering the mapping described in~\eqref{e:mat_fast},
for any choice of $\bm{\alpha}$,
$\mathbf{T}^{(\bm{\alpha})}_8$ 
satisfies
the Feig-Winograd factorization.
Thus,
all
matrices 
in this particular subspace
possess
the same 
general
fast algorithm structure,
which is shown in
Fig.~\ref{f:fw_fast}.

\subsection{Inverse Transformation}

Assuming the existence of the inverse of~$\mathbf{T}^{(\bm{\alpha})}_8$,
by direct computation,
we obtain the following expression:
\begin{align*}
(\mathbf{T}^{(\bm{\alpha})}_8)^{-1}
&=
(\mathbf{B}^{(3)}_8)^{-1} \cdot 
(\mathbf{B}^{(2)}_8)^{-1} \cdot 
(\mathbf{B}^{(1)}_8)^{-1} \cdot
(\mathbf{K}^{(\bm{\alpha})}_8)^{-1} \cdot
\mathbf{P}_8^{-1}
.
\end{align*}
However,
we notice that the following relationships hold true:

\begin{align*}
\mathbf{P}_8^{-1} 
&
=
\mathbf{P}_8^\top
,
\quad
(\mathbf{B}^{(1)}_8)^{-1} 
=
\operatorname{diag}
\left(
\frac{1}{2}\cdot \mathbf{I}_2, \mathbf{I}_6
\right)
\cdot
\mathbf{B}^{(1)}_8
,
\\
(\mathbf{B}^{(2)}_8)^{-1} 
&
=
\operatorname{diag}
\left(
\frac{1}{2}
\cdot\mathbf{I}_4, \mathbf{I}_4
\right)
\cdot
\mathbf{B}^{(2)}_8
,
\quad
(\mathbf{B}^{(3)}_8)^{-1} 
= 
\frac{1}{2}
\cdot
\mathbf{B}^{(3)}_8
.
\end{align*}
Thus,
straightforward matrix manipulation
yields:
\begin{align}
\label{equation-inverse-of-T-alpha-prime}
(\mathbf{T}^{(\bm{\alpha})}_8)^{-1}
&=
\mathbf{B}^{(3)}_8 \cdot 
\mathbf{B}^{(2)}_8 \cdot 
\mathbf{B}^{(1)}_8
\cdot
(\mathbf{K}^{(\bm{\alpha})}_8)^{-1}
\cdot
\mathbf{P}_8^{\top}
\cdot
\mathbf{D}^{(0)}_8
,
\end{align}
where
$\mathbf{D}^{(0)}_8 = \operatorname{diag}\left(
\frac{1}{8}, \frac{1}{2}, \frac{1}{4}, \frac{1}{2}, 
\frac{1}{8}, \frac{1}{2}, \frac{1}{4}, \frac{1}{2}
\right)$.

Matrices~$\mathbf{B}^{(1)}_8$, $\mathbf{B}^{(2)}_8$, and $\mathbf{B}^{(3)}_8$
represent butterfly operations.
Matrix~$\mathbf{P}_8^\top$ is a simple permutation.
Being~$\mathbf{K}^{(\bm{\alpha})}_8$ a block diagonal matrix,
its inverse is also a block diagonal.
In fact,
$(\mathbf{K}^{(\bm{\alpha})}_8)^{-1}$
is closely related to 
$\mathbf{K}^{(\bm{\alpha})}_8$;
both possessing a very similar structure.
By means of symbolic computation~\cite{MATLAB2013},
we obtain that:
\begin{align}
\label{equation-K-alpha-primo}
(\mathbf{K}^{(\bm{\alpha})}_8)^{-1}
=
(\mathbf{K}^{(\bm{\alpha'})}_8)^{\top}
,
\end{align}
where $\mathbf{K}^{(\bm{\alpha'})}_8$ 
has the structure shown in~\eqref{e:K_alpha}
with parameter vector
\begin{align}
\label{E:alphalinha}
\bm{\alpha'}=
\begin{bmatrix}\alpha_0' & \alpha_1' & \alpha_2' & \alpha_3' & \alpha_4' & \alpha_5' & \alpha_6' \end{bmatrix}^\top
\end{align} 
given by
\begin{align}
\label{equation-set}
\begin{split}
\alpha_0'&= 
[
\alpha_0\,{\alpha_6}^{2}+\left( {\alpha_2}^{2}-{\alpha_4}^{2}\right) \,\alpha_6+2\,\alpha_0\,\alpha_2\,\alpha_4+{\alpha_0}^{3}
]/\lambda
, \\
\alpha_1'&= 
{\alpha_1}
/[\alpha_1^2+\alpha_5^2 ], \\
\alpha_2'&= 
[
\alpha_2\,{\alpha_4}^{2}+\left( {\alpha_0}^{2}-{\alpha_6}^{2}\right) \,\alpha_4+2\,\alpha_0\,\alpha_2\,\alpha_6+{\alpha_2}^{3}
]/\lambda
,\\
\alpha_3'&= 1/\alpha_3, \\
\alpha_4'&= 
[
\alpha_4\,{\alpha_2}^{2}+\left( {\alpha_0}^{2}-{\alpha_6}^{2}\right) \,\alpha_2-2\,\alpha_0\,\alpha_4\,\alpha_6+{\alpha_4}^{3}
]/\lambda
, \\
\alpha_5'&= \alpha_5/ [\alpha_1^2+\alpha_5^2], \\
\alpha_6'&= 
[
\alpha_6\,{\alpha_0}^{2}+\left( {\alpha_2}^{2}-{\alpha_4}^{2}\right) \,\alpha_0-2\,\alpha_2\,\alpha_4\,\alpha_6+{\alpha_6}^{3}
]/\lambda
,
\end{split}
\end{align}
and
$\lambda =
\left({\alpha_0}^{2}+{\alpha_6}^{2}\right)^2
+\left({\alpha_2}^{2}+{\alpha_4}^{2}\right)^2
+4\,
\left(
\alpha_0\alpha_2-\alpha_4\alpha_6
\right)\left(
\alpha_2\alpha_6+\alpha_0\alpha_4
\right)$.

The inverse of $\mathbf{T}^{(\bm{\alpha})}_8$
does exist as long as
$(\mathbf{K}^{(\bm{\alpha})}_8)^{-1}$
is also well-defined.
From the set of equations~\eqref{equation-set}, 
the 
following conditions are necessary for the existence of
$(\mathbf{K}^{(\bm{\alpha})}_8)^{-1}$:
\begin{enumerate}[(i)]
\item $\alpha_3 \neq 0$, 
\item ${\alpha_1}^{2}+{\alpha_5}^{2} \neq 0 $, 
\item ${\alpha_0}^{2}+{\alpha_2}^{2}+{\alpha_4}^{2}+{\alpha_6}^{2} \neq 0 $.
\end{enumerate}

Applying~\eqref{equation-K-alpha-primo} 
in~\eqref{equation-inverse-of-T-alpha-prime} and using~\eqref{e:mat_fast},
we directly obtain that:
\begin{align}
\label{eq:inversa_is_fw}
(\mathbf{T}^{(\bm{\alpha})}_8)^{-\top}
&=
\mathbf{D}^{(0)}_8
\cdot
\mathbf{P}_8
\cdot
\mathbf{K}^{(\bm{\alpha}')}_8
\cdot 
\mathbf{B}^{(1)}_8 \cdot 
\mathbf{B}^{(2)}_8 \cdot 
\mathbf{B}^{(3)}_8\nonumber\\
&=\mathbf{P}_8
\cdot
\mathbf{D}^{(1)}_8 
\cdot
\mathbf{K}^{(\bm{\alpha'})}_8
\cdot 
\mathbf{B}^{(1)}_8 \cdot \mathbf{B}^{(2)}_8 \cdot \mathbf{B}^{(3)}_8
,
\end{align}
where
$\mathbf{D}^{(1)}_8 = \operatorname{diag}\left(
\frac{1}{8}, \frac{1}{8} , \frac{1}{4}, \frac{1}{4}, \frac{1}{2}, 
\frac{1}{2}, \frac{1}{2}, \frac{1}{2}
\right)$
and
$(\mathbf{T}^{(\bm{\alpha})}_8)^{-\top}$
is the transpose inverse of $\mathbf{T}^{(\bm{\alpha})}_8$.
However,
notice that
$\mathbf{D}^{(1)}_8\cdot\mathbf{K}^{(\bm{\alpha'})}_8$
is shaped as~\eqref{e:K_alpha}.
As a consequence, 
from~\eqref{eq:inversa_is_fw}
we can conclude that
the fast algorithm for 
the inverse transformation
is obtained by
simply
replacing  $\bm{\alpha}$ of the direct transform 
by $\bm{\alpha'}$ 
(as described in~\eqref{E:alphalinha})
and 
then
applying the transposition operation.
Thus, 
the hardware implementation of the inverse transformation
is facilitated.

\subsection{Particular Known Matrices}\label{cases}

Based on different values of the parameter vector 
$\bm{\alpha}$,
possibly distinct
Feig-Winograd approximation matrices 
can be obtained. 
In this subsection,
we furnish 
a list
of
several  
transforms 
archived in literature,
which
are particular cases
encompassed in the
proposed
Feig-Winograd formalism.
Various well-known approximations of DCT matrix
$\mathbf{C}_8$
are among these identified cases,
as shown below:

\subsubsection{Exact DCT}
Notice that for
$\bm{\alpha}_0 = \begin{bmatrix} \gamma_1 & \gamma_2 & \cdots & \gamma_7\end{bmatrix}^\top$,
we have that
$\operatorname{FW}(\bm{\alpha}_0/2) = \mathbf{C}_8$.

\subsubsection{Signed DCT}
Let
$\operatorname{sign}(\cdot)$ 
be the signum function~\cite{haweel2001new}.
For
$\bm{\alpha}_1 = \operatorname{sign}(\bm{\alpha}_0)=\begin{bmatrix}1 & 1 & 1 & 1 & 1 & 1 & 1 \end{bmatrix}^\top$,
matrix
$\operatorname{FW}(\bm{\alpha}_1)$ 
is the low-complexity matrix
associated to the $8$-point SDCT~\cite{haweel2001new}.

\subsubsection{Level 1 approximation}
For
$\bm{\alpha}_2 = \begin{bmatrix}
1 & 1 & 1 & 1 & 1 & 1/2 & 0
\end{bmatrix}^\top$, 
according to~\eqref{e:mat_fast}
we have that
$ \operatorname{FW}(\bm{\alpha}_2)$ 
is the low-complexity matrix
of
the level~1 approximation by 
Lengwehasatit-Ortega~\cite{lengwehasatit2004scalable}.

\subsubsection{Rounded DCT}
Let
$\operatorname{round}(\cdot)$ 
be the round-off function as implemented in 
C or Matlab language~\cite{MATLAB2013}.
Considering 
$\bm{\alpha}_3 = \operatorname{round}(\bm{\alpha}_0)=
\begin{bmatrix}1 & 1 & 1 & 1 & 1 & 0 & 0 \end{bmatrix}^\top$,
$\operatorname{FW}(\bm{\alpha}_3)$ 
is the approximate $8$-point DCT introduced in~\cite{cb2011}.

\subsubsection{Modified rounded DCT}
If
$\bm{\alpha}_4 = \begin{bmatrix}
1 & 1 & 0 & 1 & 0 & 0 & 0
\end{bmatrix}^\top$,
then
$ \operatorname{FW}(\bm{\alpha}_4)$ 
results in
the modified rounded $8$-point DCT introduced in~\cite{bc2012}. 

\subsubsection{DCT approximation for RF imaging}
If
$\bm{\alpha}_5 = \begin{bmatrix}
2 & 2 & 1 & 1 & 1 & 1 & 0
\end{bmatrix}^\top$,
then
$ \operatorname{FW}(\bm{\alpha}_5)$ 
is the
multiplier-free $8$-point DCT approximation for RF imaging
proposed in~\cite{multibeam2012}.

\subsubsection{Haar matrix}
Let
$\mathbf{H}_8$ 
be
the non-normalized Haar matrix of order~8~\cite[p.~159]{Jain1989}.
For
$\bm{\alpha}_6 = 
\begin{bmatrix} 0 & 0 & 0 & 1 & 1 & 1 & 0 \end{bmatrix}^\top$,
we have that
$\mathbf{H}_8 = 
\mathbf{P}^{(1)}_8\cdot
\operatorname{FW}(\bm{\alpha}_6)
\cdot
\mathbf{P}^{(2)}_8 
$,
where 
$\mathbf{P}^{(1)}_8=(1)(8\ 2\ 5\ 6\ 4\ 3\ 7)$, 
$\mathbf{P}^{(2)}_8=(1)(8\ 2\ 6)(5\ 3\ 7)(4)$ 
are permutation matrices denoted 
in cyclic notation~\cite[p.~64]{Herstein2006}.

\subsubsection{8-point transform employed in AVC/H.264}
If 
$\bm{\alpha}_7 = \begin{bmatrix}
12 & 8 & 10 & 8 & 6 & 4 & 3
\end{bmatrix}^\top$,
then
$ \operatorname{FW}(\bm{\alpha}_7)$ 
is the integer DCT employed in AVC/H.264~\cite{H264_8transform}. 

\subsubsection{8-point transform employed in HEVC/H.265}
If 
$\bm{\alpha}_8 = \begin{bmatrix}
89 & 83 & 75 & 64 & 50 & 36 & 18
\end{bmatrix}^\top$,
then
$\operatorname{FW}(\bm{\alpha}_8)$ 
is the 8-point transform employed used in HEVC/H.265~\cite{hevc1,Fuldseth2011,sullivan2012}.

\section{Approximation Criteria in the Feig-Winograd Matrix Space}
\label{section-approximation}

We aim at investigating conditions
under which
$\mathbf{T}^{(\bm{\alpha})}_8$
could
provide adequate approximations
for the $8$-point DCT.
To guide our search 
we  adopt the following 
general criteria related
to $\mathbf{T}^{(\bm{\alpha})}_8$:
\begin{enumerate}%

\item
it 
must
possess low computational complexity;

\item
it must satisfy
orthogonality 
or
nearly orthogonality~\cite{Flury1986};

\item
its inverse transformation
must
possess low computational complexity;

\item
it must
be a close approximation to the exact
DCT matrix
according to meaningful proximity measures.

\end{enumerate}

\subsection{Computational Complexity}

The computational complexity of 
the Feig-Winograd structure is essentially
quantified by its arithmetic complexity,
which is furnished by
the number of multiplications,
additions, 
and
bit-shifting operations
required
for its calculation~\cite{Heideman1988,Briggs1995}.
The multiplicative count is due to the elements
$\alpha_k$, $k=0,1,\ldots,6$ in matrix~$\mathbf{K}^{(\bm{\alpha})}_8$.
However,
for judiciously selected values of~$\bm{\alpha}$,
the resulting multiplicative count can be lower.
One possibility is to restrict 
the
elements
$\alpha_k$, $k=0,1,\ldots,6$
to the set of dyadic rationals~\cite{britanak2007discrete}.

Although this representation may lead to good approximations,
it implies an increase
in the additive complexity as well as in the number
of required bit-shifting operations~\cite{britanak2007discrete,Cheung1993galileo}.
A more effective approach is
to consider
only zero adder representation quantities~\cite{britanak2007discrete}.
In other words,
numbers whose binary representation
requires no extra adders.
This is true for powers of two.
Aiming at the full  
computational complexity minimization,
we further restricted the choice of
$\alpha_k$, $k=0,1,\ldots,6$,
to the following set of numbers:
$\mathcal{P}=\{0, \pm1/2, \pm1, \pm2 \}$.
In terms 
of digital arithmetic circuits
the multiplication by such elements
requires no
additions and only minimal bit-shifting operations.
This implies null multiplicative complexity in
considered DCT approximations.

Over the set $\mathcal{P}$,
the worst case scenario
in terms of computational complexity
is to select
non-null parameters in $\{\pm1/2, \pm2\}$.
This would imply 28~additions and 22~bit-shifting operations.
Considering the already identified
DCT approximations in the Feig-Winograd matrix space,
the expected complexity for good approximations
may be typically lower.
For example,
the $8$-point SDCT and the rounded
DCT---both in the Feig-Winograd matrix subspace---require
24/0 and 22/0 additions/bit-shifting operations, 
respectively~\cite{haweel2001new,cb2011}.

Expressions
\eqref{e:mat_fast} and~\eqref{e:K_alpha}
as well as 
Fig.~\ref{f:fw_fast}
suggests that the
additive complexity of the Feig-Winograd
transformations
is equal to 14
plus the additions confined 
in matrix $\mathbf{K}^{(\bm{\alpha})}_8$.
Similarly,
the number of bit-shifting operations
is fully determined
by the elements of $\mathbf{K}^{(\bm{\alpha})}_8$.
Let
$\theta(\cdot)$
and 
$\phi(\cdot)$
be the functions that return, respectively,
the number of non-null elements 
and 
the number of elements in $\{\pm1/2,\pm2\}$ 
of their vector arguments.
For $\bm{\alpha}\in \mathcal{P}^7$,
the addition and bit-shifting counts,
respectively denoted by
$\mathcal{A}(\bm{\alpha})$
and
$\mathcal{S}(\bm{\alpha})$
are given by:
\begin{align}
\label{eq:additions}
\mathcal{A}(\bm{\alpha}) 
=&
14 +
2 \cdot 
\max
\left\{
1,\theta\left(
\begin{bmatrix}\alpha_1 & \alpha_5\end{bmatrix}^\top
\right) 
\right\} 
+ 
4 \cdot 
\max
\left\{
1,\theta\left(
\begin{bmatrix}\alpha_0 & \alpha_2 & \alpha_4 & \alpha_6\end{bmatrix}^\top
\right)
\right\}
-
6
,
\\
\label{eq:bit-shifting}
\mathcal{S}(\bm{\alpha})
=& 
2 \cdot 
\phi\left(
\begin{bmatrix}\alpha_3\end{bmatrix}
\right) 
+
2 \cdot 
\phi\left(
\begin{bmatrix}\alpha_1 & \alpha_5\end{bmatrix}^\top
\right) 
+ 
4 \cdot 
\phi\left(
\begin{bmatrix}\alpha_0 & \alpha_2 & \alpha_4 & \alpha_6\end{bmatrix}^\top
\right)
,
\end{align}
By inspecting above expressions,
we notice
$14 \leq \mathcal{A}(\bm{\alpha}) \leq 28$
and
$0 \leq \mathcal{S}(\bm{\alpha})\leq 22$,
for 
$\bm{\alpha}\in \mathcal{P}^7$.
Thus,
the theoretical lower-bound for the complexity of
the Feig-Winograd matrices
is 14 additions.
Table~\ref{t:comp} summarizes the operation counts discussed above.

\begin{table}
\centering
\caption{Arithmetic complexity of the Feig-Winograd fast algorithm
according to the employed number representation}
\label{t:comp}
\begin{tabular}{lccc}
\toprule
Number representation         & Mult.       & Add.     & Bit-shifting  \\
\midrule
Float point     & 22          & 28            & 0             \\
8-bit dyadic rationals & 0     & 42    & 21            \\
Zero adder rationals
                & 0           & at most 28    & at most 22    \\
\bottomrule
\end{tabular}
\end{table}

\subsection{Orthogonality or Near Orthogonality}

Orthogonality is often a desirable property sought
in a DCT approximation matrix~\cite{britanak2007discrete,lengwehasatit2004scalable}.
Among several 
orthogonalization methods~\cite{Watkins2004,higham2008functions},
we separate the one based on 
the polar decomposition~\cite{Higham1986,Higham1988}.
To
orthogonalize $\mathbf{T}^{(\bm{\alpha})}_8$,
such procedure
requires only one matrix 
given by~\cite{cintra2011integer}:
\begin{align*}
\mathbf{S}_8
&
=
\sqrt{
[\mathbf{T}^{(\bm{\alpha})}_8 
\cdot 
(\mathbf{T}^{(\bm{\alpha})}_8)^{\top}
]^{-1}
}
,
\end{align*}
where
$\sqrt{\cdot}$ denotes
the matrix
square root operation~\cite{Higham1987,MATLAB2013}.
The resulting orthogonal DCT approximation
is furnished 
by~\cite{cintra2011integer,cb2011,bc2012,multibeam2012}:
\begin{align}
\label{equation--orthogonal-approximation}
\hat{\mathbf{C}}_8 = \mathbf{S}_8 \cdot \mathbf{T}^{(\bm{\alpha})}_8
.
\end{align}
Importantly, 
this method preserves the structure and low-complexity
of~$\mathbf{T}^{(\bm{\alpha})}_8$~\cite{Higham2004,bas2008,lengwehasatit2004scalable,cb2011,bas2009} 
and allow lossless transformation.

In the context of image compression,
if $\mathbf{S}_8$ is a diagonal matrix,
then it does not
introduce any additional computational overhead.
In this case,
matrix~$\mathbf{S}_8$ 
can be merged into the quantization step
of JPEG-like compression schemes~\cite{lengwehasatit2004scalable, bas2008, bas2009, bas2011, cb2011, bc2012, bayer201216pt}.

For matrix~$\mathbf{S}_8$ to be diagonal,
it is sufficient
that~\cite{cintra2011integer}:
\begin{align}
\label{equation-condition-1}
\mathbf{T}^{(\bm{\alpha})}_8
\cdot
(\mathbf{T}^{(\bm{\alpha})}_8)^\top
=
[\text{diagonal matrix}]
.
\end{align}
By explicitly 
calculating~$\mathbf{T}^{(\bm{\alpha})}_8 \cdot (\mathbf{T}^{(\bm{\alpha})}_8)^\top$ 
and using symbolic computation~\cite{MATLAB2013},
we obtain that
a sufficient condition for~\eqref{equation-condition-1}
to hold true
is:
\begin{align}\label{E:ortho2}
\alpha_0\cdot
(\alpha_2 - \alpha_4)
=
\alpha_6\cdot
(\alpha_2
+ 
\alpha_4)
.
\end{align}

If $\mathbf{T}^{(\bm{\alpha})}_8$ 
satisfies \eqref{equation-condition-1} or \eqref{E:ortho2},
then we have that
\begin{align*}
\mathbf{S}_8
=
\operatorname{diag}
(
s_0,s_1,s_2,s_1,s_0,s_1,s_2,s_1
)
,
\end{align*}
where 
$s_0 = 1/(2^{3/2} \alpha_3)$,
$s_1 = 1/\sqrt{2(\alpha_6^2+ \alpha_4^2+ \alpha_2^2+ \alpha_0^2)}$,
and
$s_2 = 1/(2 \sqrt{\alpha_5^2+\alpha_1^2})$.
If matrix~$\mathbf{T}^{(\bm{\alpha})}_8$
does not satisfy~\eqref{equation-condition-1},
but 
$\mathbf{T}^{(\bm{\alpha})}_8 
\cdot 
(\mathbf{T}^{(\bm{\alpha})}_8)^{\top}$
is a nearly diagonal matrix~\cite{Flury1986},
then
$\mathbf{S}_8$
is also nearly diagonal.
Thus,
the following approximation
for the orthogonalization matrix
can be taken into consideration:
\begin{align}
\label{equation-S-hat}
\hat{\mathbf{S}}_8
= 
\sqrt{
\big\{
\operatorname{diag}
[\mathbf{T}^{(\bm{\alpha})}_8 
\cdot 
(\mathbf{T}^{(\bm{\alpha})}_8)^{\top}
]
\big\}^{-1}
}
,
\end{align}
where $\operatorname{diag}(\cdot)$
returns a diagonal matrix 
with the same diagonal elements
of its matrix argument~\cite[p.~2]{seber2008matrix};
i.e., $\operatorname{diag}(\cdot)$
operates consistently with Matlab usage~\cite{MATLAB2013}.
Consequently,
the obtained nearly orthogonal
approximation
is
given by
\begin{align}
\label{equation-nearly-orthogonal-approximation}
\hat{\mathbf{C}}_8 = 
\hat{\mathbf{S}}_8
\cdot 
\mathbf{T}^{(\bm{\alpha})}_8
.
\end{align}

To quantify how close a matrix is to the diagonal form,
we adopt the deviation from diagonality measure~\cite{Flury1986},
which is described as follows.
Let~$\mathbf{M}$
be a square matrix.
The deviation from diagonality measure of~$\mathbf{M}$
is given by:
\begin{align}
\label{equation-dfd}
\operatorname{\delta}(\mathbf{M})
=
1 -
\frac{\|\operatorname{diag}(\mathbf{M}) \|_\text{F}^2}
{\| \mathbf{M}\|_\text{F}^2}
,
\end{align}
where $\|\cdot\|_\text{F}$ denotes 
the Frobenius norm~\cite[p.~115]{Watkins2004}.
For diagonal matrices, function $\delta(\cdot)$ returns zero.
Both the $8$-point SDCT~\cite{haweel2001new}
and the BAS approximation proposed in~\cite{Bouguezel2008}
are nonorthogonal
and
good DCT approximations.
Their orthogonalization matrices
have
deviation from diagonality 
equal to
0.20 and 0.1774,
respectively.
Thus we adopt these particular measurements 
as reference values
for identifying 
nearly diagonal 
orthogonalization
matrices
in the context of DCT approximation.

\subsection{Structure and Complexity of Inverse Transformation}

Not only it is important to identify
low-complexity approximations
but also to guarantee that 
the 
associated
inverse transformations
also possess low computational complexity.

For orthogonalizable approximations, 
the following holds true:
$\hat{\mathbf{C}}_8^{-1}
=
\hat{\mathbf{C}}_8^\top
=
(\mathbf{T}^{(\bm{\alpha})}_8)^\top
\cdot
\mathbf{S}_8^\top
$.
Therefore, 
in this case,
the inverse transformation inherits
the low-complexity properties of
$\mathbf{T}^{(\bm{\alpha})}_8$.
Moreover,
for image compression purposes,
matrix $\mathbf{S}_8^\top$
can also be absorbed in 
quantization step.

For the nonorthogonal case,
let us assume that~$\mathbf{T}^{(\bm{\alpha})}_8$
is a low-complexity transformation.
The set of equations~\eqref{equation-set}
furnishes
closed-form expressions
for the multiplicative elements
$\alpha'_k$, $k=0,1,\ldots,6,$
of
$(\mathbf{T}^{(\bm{\alpha})}_8)^{-1}$.
Thus,
for the inverse transformation 
to possess null multiplicative
complexity,
it is sufficient to ensure that
$\alpha'_k\in\mathcal{P}$,
for $k=0,1,\ldots,6$.

As an illustration,
we consider 
the 
$8$-point SDCT,
which is
a nonorthogonal transform.
The SDCT low-complexity matrix
is
defined
on the Feig-Winograd space
with parameter vector given by
$\bm{\alpha}_1 = \begin{bmatrix}1 & 1 & 1 & 1 & 1 & 1 & 1 \end{bmatrix}^\top$.
We notice that
$\operatorname{FW}(\bm{\alpha}_1)
\cdot
\operatorname{FW}(\bm{\alpha}_1')
= 
8 \cdot \mathbf{I}_8
$,
where
$\bm{\alpha}_1' = \begin{bmatrix}2 & 1 & 2 & 1 & 0 & 1 & 0 \end{bmatrix}^\top$
satisfies
the set of equations~\eqref{equation-set}.

\subsection{Proximity Measures}

In order to evaluate 
candidate DCT approximations,
we consider the 
following figures of merit:
(i)~the total error energy~\cite{cb2011};
(ii)~the mean square error (MSE)~\cite{Wang2009};
(iii)~the 
unified coding gain~\cite{Katto1991,britanak2007discrete,Goyal2001};
and
(iv)~the transform efficiency~\cite{britanak2007discrete}.
The total error energy and MSE
are employed to quantify
how close
a given DCT approximation~$\hat{\mathbf{C}}_N$
is to the exact DCT matrix~$\mathbf{C}_N$.
The coding gain and transform efficiency
capture
the coding performance
of a given transformation~\cite{britanak2007discrete}.

For coding performance evaluation,
we assume 
that the data are
modeled as
a first-order
Gaussian Markov process
with zero-mean, unit variance, and correlation coefficient
$\rho=0.95$~\cite{Liang2001,britanak2007discrete,Katto1991}.
Then, 
the
$(m,n)$-th element of
the covariance matrix~$\mathbf{R}_\mathbf{x}$
of the input signal $\mathbf{x}$
is given by
$r^{(\mathbf{x})}_{m,n} = \rho^{|m-n|}$~\cite{britanak2007discrete}.
Natural images satisfy 
above assumptions~\cite{Liang2001}.
Below
we briefly describe the selected figures of merit.

\subsubsection{Total Error Energy}

Each row of a given transform matrix can
be understood as finite impulse response filter
with associate transfer 
functions~\cite{cintra2011integer,Strang1999}.
Based on~\cite{haweel2001new},
the magnitude of the difference between
the transfer functions of the DCT and the SDCT
was advanced in~\cite{cb2011} 
as a similarity measure.
Such measure,
termed total error energy,
was further employed 
as a proximity measure
between several other
approximations~\cite{cb2011,bayer201216pt}.
Although originally defined
on the spectral domain~\cite{haweel2001new},
the total error energy
can be given a simple 
matrix form by means of the 
Parseval theorem~\cite[p.~18]{Jain1989}.
The total error energy $\epsilon$ for a given
DCT approximation matrix~$\hat{\mathbf{C}}_N$
is furnished by
$\epsilon=
\pi
\cdot
\|
\mathbf{C}_N
-
\hat{\mathbf{C}}_N
\|_\text{F}^2
$.

\subsubsection{Mean Square Error}

The mean square error (MSE) 
for an approximation matrix $\hat{\mathbf{C}}_N$
is defined as~\cite{Liang2001,britanak2007discrete}
\begin{align*}
\operatorname{MSE} 
=
\frac{1}{N}
\cdot
\operatorname{tr}
\left(
(\mathbf{C}_N - \hat{\mathbf{C}}_N)
\cdot 
\mathbf{R}_\mathbf{x}
\cdot 
(\mathbf{C}_N - \hat{\mathbf{C}}_N)^{\top}
\right)
, 
\end{align*}
where $\operatorname{tr}(\cdot)$ is the trace function~\cite{Merikoski1984}.

To maintain the compatibility 
between the approximation and the \textit{exact} DCT outputs,
the MSE between the DCT and the approximation coefficients
should be minimized~\cite{Liang2001,britanak2007discrete}.

\subsubsection{Unified Transform Coding Gain}

We adopt the unified coding gain,
which generalizes the usual coding gain~\cite{Katto1991}.
Let $\mathbf{h}_k$ and $\mathbf{g}_k$ be the $k$th row of 
$\hat{\mathbf{C}}_N$ and $\hat{\mathbf{C}}_N^{-1}$,
respectively.
Thus,
the coding gain of~$\hat{\mathbf{C}}_N$
is given  by:
\begin{align*}
C_g
=10\cdot\log_{10}\!\left[
\prod_{k=1}^N
\frac{1}{(A_k \cdot B_k)^{1/N}}\right]
\quad
(\text{in dB})
,
\end{align*}
where
$
A_k
=
\operatorname{su}
[(
\mathbf{h}_k^\top
\cdot
\mathbf{h}_k
)
\circ
\mathbf{R}_\mathbf{x}]
$,
$\operatorname{su}(\cdot)$
returns the sum of the elements of its matrix 
argument~\cite{Merikoski1984},
operator $\circ$ is the element-wise matrix product~\cite{seber2008matrix},
$
B_k = \| \mathbf{g}_k \|_2^2
$,
and
$\|\cdot\|_2$ is the Euclidean norm.

Approximations exhibiting high transform coding gains 
can 
compact
more energy into 
few
coefficients~\cite{Liang2001}.
The transform coding gain for the KLT and $8$-point DCT
are
8.8462
and
8.8259,
respectively~\cite{britanak2007discrete}.

\subsubsection{Transform Efficiency}

Let $r^{(\mathbf{X})}_{m,n}$
be the $(m,n)$-th entry
of the covariance matrix of the transformed signal~$\mathbf{X}$,
which is given by
$\mathbf{R}_\mathbf{X} = \hat{\mathbf{C}}_N \cdot \mathbf{R}_\mathbf{x} \cdot \hat{\mathbf{C}}_N^{\top}$.
The
transform efficiency is defined as~\cite{Cham1989,britanak2007discrete}
\begin{align*}
\eta
=
\frac{
\sum_{m=1}^N
|r^{(\mathbf{X})}_{m,m}|
}
{
\sum_{m=1}^N \sum_{n=1}^N 
|r^{(\mathbf{X})}_{m,n}|
}
\cdot 100
.
\end{align*}
The transform efficiency $\eta$
measures the decorrelation ability 
of the transform~\cite{britanak2007discrete}.
The optimal KLT converts signals into 
completely uncorrelated coefficients
and has a 
transform efficiency equal to 100,
for any value of $\rho$.

\section{Optimization over the Feig-Winograd Space 
and new transformations}
\label{S:results}

\subsection{Multicriteria Optimization}

In this section,
we 
propose
and solve
an optimization problem
considering the 
discussed 
criteria
in the previous section.
More formally,
we 
aim at solving the following
multicriteria optimization problem~\cite{Ehrgott2000,Miettinen1999}:
\begin{align}
\label{eq:multicriteria}
\arg\min_{\bm{\alpha}}
\left(
\epsilon(\bm{\alpha}),
\operatorname{MSE}(\bm{\alpha}),
 -C_g(\bm{\alpha}),
-\eta(\bm{\alpha}),
\mathcal{A}(\bm{\alpha}),
\mathcal{S}(\bm{\alpha})
\right)
.
\end{align}
In~\eqref{eq:multicriteria}, 
the dependence of the proximity measures on the parameter vector $\bm{\alpha}$ is emphasized. 
Since
$C_g(\bm{\alpha})$ 
and 
$\eta(\bm{\alpha})$
are to be maximized,
we
considered them in negative form.
We emphasize that all the objective functions are equally relevant 
and
therefore no ranking over themselves is considered. 
This procedure is widely used in different optimization problems that consider several objective functions~\cite{Oliveira2010}.

To
address above optimization problem,
we must identify
the search space,
the set of constraints,
and
the solving method.
As
we
require 
the candidate solutions~$\bm{\alpha}$
to generate
low-complexity matrices~$\operatorname{FW}(\bm{\alpha})$,
we have that $\alpha_k \in \mathcal{P}$,
$k = 0,1,\ldots,6$.
Thus
the search space for~\eqref{eq:multicriteria}
is the set $\mathcal{P}^7$.
Additionally,
we notice that~\eqref{eq:multicriteria}
is a constrained problem.
In fact,
we only consider
candidate solutions
whose inverse transform 
possesses low arithmetic complexity.
In other words,
both direct and inverse transformations
require only numerical values defined on 
the set~$\mathcal{P}=\{0, \pm1/2, \pm1, \pm2 \}$.

In view of the above restriction,
we recognize that~\eqref{eq:multicriteria}
is not an analytically tractable problem.
However,
because
the search space~$\mathcal{P}^7$
contains only
$7^7=823543$ elements,
exhaustive
computational search is practical.
Exhaustive search is guaranteed to find solutions
and
could indeed identify the \emph{efficient} solutions
of~\eqref{eq:multicriteria}~\cite[p.~24]{Ehrgott2000}.
Let
$\mathcal{F}=
\left\{
\epsilon(\cdot),
\operatorname{MSE}(\cdot),
-C_g(\cdot),
-\eta(\cdot),
\mathcal{A}(\cdot),
\mathcal{S}(\cdot)
\right\}$ 
be the set objective functions considered in~\eqref{eq:multicriteria}.
Each of the six values given by these objective functions provide a vector in $\mathbb{R}^6$. Since there is no canonical order in $\mathbb{R}^6$, is considered the Pareto ordering to find the values $\bm{\alpha}$ that are solutions of~\eqref{eq:multicriteria}, which define the set of efficient solutions~\cite{Ehrgott2000}:
\begin{align*}
\Big\{
\bm{\alpha}^\ast\in\mathcal{P}^7 
\colon
\text{%
there is no
$\bm{\alpha}\in\mathcal{P}^7$
such that }
\text{%
$f(\bm{\alpha})\leq f(\bm{\alpha}^\ast)$
for all
$f\in\mathcal{F}$
and }
\text{%
$f_0(\bm{\alpha})<f_0(\bm{\alpha}^\ast)$
for some
$f_0\in\mathcal{F}$}
\Big\}
.
\end{align*}

\subsection{Efficient Solutions and New Transformations}

As a result of the computational search,
we obtained 16~distinct
efficient solutions
referred to as $\bm{\alpha}^\ast_i$,
$i = 1,2,\ldots,16$,
as shown in 
Table~\ref{table-solutions}.
Each efficient solution
implies an
approximate DCT matrix
given by~$\bm{\mathsf{T}}^{(i)}_8 = \operatorname{FW}(\bm{\alpha}^\ast_i)$.
The explicit form 
of 
$\bm{\mathsf{T}}^{(i)}_8$,
$i = 1,2,\ldots,16$,
is directly obtained from~\eqref{e:mat_fast}.

\begin{table}
\centering

\caption{Efficients solutions in the Feig-Winograd space}
\label{table-solutions}
\begin{tabular}{cc}
\toprule
$i$ & 
Efficient solution ($\bm{\alpha}^\ast_i$) 
\\
\midrule  
1 & 
$\begin{bmatrix} 1 & 1 & 1 & 1 & 1 & 1/2 & 0 \end{bmatrix}^\top$
\\
2 & 
$\begin{bmatrix} 1 & 1 & 1 & 1 & 1 & 0 & 0 \end{bmatrix}^\top$ 
\\
3 &
$\begin{bmatrix} 1 & 1 & 0 & 1 & 0 & 0 & 0 \end{bmatrix}^\top$ 
\\
4 &
$\begin{bmatrix} 1 & 2 & 0 & 1 & 0 & 1 & 0 \end{bmatrix}^\top$ 
\\
5 & 
$\begin{bmatrix} 0 & 1 & 1 & 1 & 1 & 0 & 0 \end{bmatrix}^\top$
\\
6 & 
$\begin{bmatrix} 0 & 2 & 1 & 1 & 1 & 1 & 0 \end{bmatrix}^\top$
\\
7 & 
$\begin{bmatrix} 0 & 2 & 2 & 1 & 1 & 1 & 0 \end{bmatrix}^\top$
\\
8 & 
$\begin{bmatrix} 2 & 2 & 0 & 1 & 0 & 1 & 1/2 \end{bmatrix}^\top$
\\
9 & 
$\begin{bmatrix} 1 & 2 & 1 & 1 & 1 & 1 & 0 \end{bmatrix}^\top$ 
\\
10 & 
$\begin{bmatrix} 1 & 1 & 0 & 1 & 0 & 1/2 & 0 \end{bmatrix}^\top$ 
\\
11 & 
$\begin{bmatrix} 0 & 1 & 1 & 1 & 1 & 1/2 & 0 \end{bmatrix}^\top$
\\
12 & 
$\begin{bmatrix} 0 & 1 & 2 & 1 & 1 & 1/2 & 0 \end{bmatrix}^\top$
\\
13 & 
$\begin{bmatrix} 0 & 2 & 1 & 1 & 1/2 & 1 & 0 \end{bmatrix}^\top$
\\
14 & 
$\begin{bmatrix} 0 & 1 & 1 & 1 & 1/2 & 1/2 & 0 \end{bmatrix}^\top$
\\
15 & 
$\begin{bmatrix} 2 & 1 & 0 & 1 & 0 & 1/2 & 1/2 \end{bmatrix}^\top$
\\
16 & 
$\begin{bmatrix} 1 & 1 & 1 & 1 & 0 & 0 & 0 \end{bmatrix}^\top$
\\
\bottomrule
\end{tabular} 

\end{table}

We notice that
among the obtained 
efficient solutions,
some of them can be linked to 
orthogonal approximations
already archived 
in literature.
In particular,
we have
that
(i)~$\bm{\mathsf{T}}^{(1)}_8$
is
the level~1 approximation suggested by 
Lengwehasatit-Ortega~\cite{lengwehasatit2004scalable},
(ii)~$\bm{\mathsf{T}}^{(2)}_8$
is the rounded DCT introduced in~\cite{cb2011},
and
(iii)~$\bm{\mathsf{T}}^{(3)}_8$ is the 
low-complexity DCT approximate proposed in~\cite{bc2012}.

On the other hand, 
matrices
$\bm{\mathsf{T}}^{(4)}_8$,
$\bm{\mathsf{T}}^{(5)}_8$,
\ldots,
$\bm{\mathsf{T}}^{(16)}_8$
are \emph{new} 
transformations.
Except for $\bm{\mathsf{T}}^{(16)}_8$,
all of them satisfy~\eqref{equation-condition-1}
and
consequently lead to orthogonal approximations.

\subsubsection{Orthogonal approximations}

A careful examination
reveals the following
relationships among
the orthogonalizable approximations:
\begin{align*}
\bm{\mathsf{T}}^{(9)}_8
&= 
\mathbf{D}^{(1)}_8
\cdot
\bm{\mathsf{T}}^{(1)}_8,
\quad
\bm{\mathsf{T}}^{(10)}_8
= 
\mathbf{D}^{(2)}_8
\cdot
\bm{\mathsf{T}}^{(4)}_8,
\quad
\bm{\mathsf{T}}^{(11)}_8
= 
\mathbf{D}^{(2)}_8
\cdot
\bm{\mathsf{T}}^{(6)}_8,
\quad
\\
\bm{\mathsf{T}}^{(12)}_8
&= 
\mathbf{D}^{(2)}_8
\cdot
\bm{\mathsf{T}}^{(7)}_8,
\quad
\bm{\mathsf{T}}^{(13)}_8
= 
\mathbf{D}^{(3)}_8
\cdot
\bm{\mathsf{T}}^{(7)}_8,
\quad
\bm{\mathsf{T}}^{(14)}_8
= 
\mathbf{D}^{(4)}_8
\cdot
\bm{\mathsf{T}}^{(7)}_8,
\\
\bm{\mathsf{T}}^{(15)}_8
&= 
\mathbf{D}^{(3)}_8
\cdot
\bm{\mathsf{T}}^{(8)}_8,
\end{align*}
where 
$\mathbf{D}^{(1)}_8= \operatorname{diag}(1, 1, 2, 1, 1, 1, 2, 1)$,
$\mathbf{D}^{(2)}_8=\operatorname{diag}
\left(1, 1,\frac{1}{2}, 1, 1, 1,\frac{1}{2}, 1\right)$, 
$\mathbf{D}^{(3)}_8=\operatorname{diag}
\left(1, \frac{1}{2}, 1, \frac{1}{2}, 1, \frac{1}{2}, 1, \frac{1}{2}\right)$,
and
$\mathbf{D}^{(4)}_8=\operatorname{diag}
\left(1, \frac{1}{2}, \frac{1}{2}, \frac{1}{2}, 1, \frac{1}{2}, \frac{1}{2}, \frac{1}{2}\right)$.

As a consequence,
although
these transformations
are pair-wise different,
the resulting
orthogonal approximations~(cf.~\eqref{equation--orthogonal-approximation})
derived from them
may not be.
This is because the orthogonalization operation
described in~\eqref{equation--orthogonal-approximation}
is not affected 
by
the left multiplication of
a diagonal matrix of positive elements. 

Therefore, 
in this sense,
each of the following sets
contains equivalent solutions~\cite[p.~332]{Hungerford1974}
with respect to the DCT approximation procedure based 
on the polar decomposition:
\begin{align}
\label{equation-equivalence-classes}
\begin{split}
\big\{
\bm{\mathsf{T}}^{(1)}_8,
\bm{\mathsf{T}}^{(9)}_8
\big\}
,
&
\big\{
\bm{\mathsf{T}}^{(4)}_8,
\bm{\mathsf{T}}^{(10)}_8
\big\}
,
\big\{
\bm{\mathsf{T}}^{(6)}_8,
\bm{\mathsf{T}}^{(11)}_8
\big\}
,
\big\{
\bm{\mathsf{T}}^{(8)}_8,
\bm{\mathsf{T}}^{(15)}_8
\big\},
\\
&
\big\{
\bm{\mathsf{T}}^{(7)}_8,
\bm{\mathsf{T}}^{(12)}_8,
\bm{\mathsf{T}}^{(13)}_8,
\bm{\mathsf{T}}^{(14)}_8
\big\}
.
\end{split}
\end{align}
Thus,
for instance,
although 
the new
matrix~$\bm{\mathsf{T}}^{(9)}_8$
is not explicitly listed in literature,
it is
equivalent
to
the
Lengwehasatit-Ortega 
approximation~$\bm{\mathsf{T}}^{(1)}_8$~\cite{lengwehasatit2004scalable}.
The remaining matrices
found no equivalence to 
any
approximation listed in literature,
being structurally new.
Therefore,
we separate
$\bm{\mathsf{T}}^{(4)}_8$,
$\bm{\mathsf{T}}^{(5)}_8$,
$\bm{\mathsf{T}}^{(6)}_8$,
$\bm{\mathsf{T}}^{(7)}_8$,
and
$\bm{\mathsf{T}}^{(8)}_8$
as 
genuinely new transformations.

\subsubsection{Nonorthogonal approximation}

Among the efficient solutions,
we obtained only one transformation that
leads to 
a nonorthogonal DCT approximation.
This particular solution, referred to as~$\bm{\mathsf{T}}^{(16)}_8$,
is a new transformation 
and
is given
by
\begin{align*}
\bm{\mathsf{T}}^{(16)}_8
=
\begin{bmatrix}
\begin{smallmatrix}
1 & \phantom{-}1 & \phantom{-}1 & \phantom{-}1 & \phantom{-}1 & \phantom{-}1 & \phantom{-}1 & \phantom{-}1 \\
1 & \phantom{-}1 & \phantom{-}0 & \phantom{-}0 & \phantom{-}0 & \phantom{-}0 & -1 & -1 \\
1 & \phantom{-}0 & \phantom{-}0 & -1 & -1 & \phantom{-}0 & \phantom{-}0 & \phantom{-}1 \\
1 & \phantom{-}0 & -1 & \phantom{-}0 & \phantom{-}0 & \phantom{-}1 & \phantom{-}0 & -1 \\
1 & -1 & -1 & \phantom{-}1 & \phantom{-}1 & -1 & -1 & \phantom{-}1 \\
0 & -1 & \phantom{-}0 & \phantom{-}1 & -1 & \phantom{-}0 & \phantom{-}1 & \phantom{-}0 \\
0 & -1 & \phantom{-}1 & \phantom{-}0 & \phantom{-}0 & \phantom{-}1 & -1 & \phantom{-}0 \\
0 & \phantom{-}0 & \phantom{-}1 & -1 & \phantom{-}1 & -1 & \phantom{-}0 & \phantom{-}0
\end{smallmatrix}
\end{bmatrix}
.
\end{align*}
Notice that
$\bm{\mathsf{T}}^{(16)}_8$
does not satisfy
conditions~\eqref{equation-condition-1}--\eqref{E:ortho2}.
Considering 
the deviation from diagonality measure 
discussed in~\cite{Flury1986} 
(cf.~\eqref{equation-dfd}),
we have
$\operatorname{\delta}
\left(
\bm{\mathsf{T}}^{(16)}_8 \cdot (\bm{\mathsf{T}}^{(16)}_8)^\top
\right)
=0.125$.
For comparison,
the well-known $8$-point SDCT
also
furnishes a near diagonal matrix~\cite{haweel2001new},
whose deviation from diagonality
is
\mbox{0.20}. 
In a sense,
matrix $\bm{\mathsf{T}}^{(16)}_8$
is
``more orthogonal''
than the SDCT low-complexity matrix.
Thus,
we accept
$\bm{\mathsf{T}}^{(16)}_8$
as a near orthogonal matrix
adequate for DCT approximation.
Consequently,
the associate 
nonorthogonal approximation
is derived from~\eqref{equation-nearly-orthogonal-approximation}
and
is
given
by
$\hat{\mathbf{S}}^{(16)}_8 \cdot \bm{\mathsf{T}}^{(16)}_8$,
where
$\hat{\mathbf{S}}^{(16)}_8
=
\mathrm{diag}
\left(
\frac{1}{\sqrt{8}},
\frac{1}{2},
\frac{1}{2},
\frac{1}{2},
\frac{1}{\sqrt{8}},
\frac{1}{2},
\frac{1}{2},
\frac{1}{2}
\right)
$
is obtained from~\eqref{equation-S-hat}.

Table~\ref{t:cases1} summarizes
above discussion 
on
the new approximations and their relationships.
Notice that all proposed approximations allow perfect reconstruction.
Most of them allow orthogonal approximation;
thus they are \emph{lossless} transforms~\cite[p.~xi]{Strang1996}.
Although $\bm{\mathsf{T}}^{(16)}_8$ is a nonorthogonal transform, 
it permits perfect reconstruction,
because it possesses a well-defined invertible matrix.

\begin{table}
\centering
\caption{Efficient Feig-Winograd DCT approximations}
\label{t:cases1}
\begin{tabular}{ccl}
\toprule
Transform & 
Orthogonalizable? &
Description 
\\
\midrule  
$\bm{\mathsf{T}}^{(1)}_8$ & 
Yes & Proposed in~\cite{lengwehasatit2004scalable}\\
$\bm{\mathsf{T}}^{(2)}_8$ & 
Yes & Proposed in~\cite{cb2011}\\
$\bm{\mathsf{T}}^{(3)}_8$ & 
Yes & Proposed in~\cite{bc2012}\\
$\bm{\mathsf{T}}^{(4)}_8$ & 
Yes & New transformation \\
$\bm{\mathsf{T}}^{(5)}_8$ & 
Yes & New transformation \\
$\bm{\mathsf{T}}^{(6)}_8$ & 
Yes & New transformation \\
$\bm{\mathsf{T}}^{(7)}_8$ & 
Yes & New transformation \\
$\bm{\mathsf{T}}^{(8)}_8$ & 
Yes & New transformation \\
$\bm{\mathsf{T}}^{(9)}_8$ & 
Yes & New, but equivalent to~$\bm{\mathsf{T}}^{(1)}_8$\\
$\bm{\mathsf{T}}^{(10)}_8$ & 
Yes & New, but equivalent to~$\bm{\mathsf{T}}^{(4)}_8$ \\
$\bm{\mathsf{T}}^{(11)}_8$ &
Yes & New, but equivalent to~$\bm{\mathsf{T}}^{(6)}_8$ \\
$\bm{\mathsf{T}}^{(12)}_8$ &
Yes & New, but equivalent to~$\bm{\mathsf{T}}^{(7)}_8$ \\
$\bm{\mathsf{T}}^{(13)}_8$ & 
Yes & New, but equivalent to~$\bm{\mathsf{T}}^{(7)}_8$ \\
$\bm{\mathsf{T}}^{(14)}_8$ & 
Yes & New, but equivalent to~$\bm{\mathsf{T}}^{(7)}_8$ \\
$\bm{\mathsf{T}}^{(15)}_8$ & 
Yes & New, but equivalent to~$\bm{\mathsf{T}}^{(8)}_8$ \\
$\bm{\mathsf{T}}^{(16)}_8$ & 
No & New transformation \\
\bottomrule
\end{tabular} 
\end{table}

\subsection{Assessment of the New Approximations}

We submitted all obtained efficient solutions
to the approximation procedure described
in~\eqref{equation--orthogonal-approximation}
and~\eqref{equation-nearly-orthogonal-approximation},
depending on whether a given solution
satisfies~\eqref{equation-condition-1} or not.
The derived approximations
were assessed by means
of
(i)~arithmetic complexity evaluation
and
(ii)~proximity measures with respect to the exact DCT.

Only non-equivalent  transformations
were considered,
as discussed in~\eqref{equation-equivalence-classes}.
We also notice
that 
equivalent transformations
possess exactly the same computational complexity.
Thus,
we only
considered
the following transformations:
$\bm{\mathsf{T}}^{(1)}_8$,
$\bm{\mathsf{T}}^{(2)}_8$,
$\bm{\mathsf{T}}^{(3)}_8$,
$\bm{\mathsf{T}}^{(4)}_8$,
$\bm{\mathsf{T}}^{(5)}_8$,
$\bm{\mathsf{T}}^{(6)}_8$,
$\bm{\mathsf{T}}^{(7)}_8$,
$\bm{\mathsf{T}}^{(8)}_8$,
and
$\bm{\mathsf{T}}^{(16)}_8$.
Table~\ref{t:cases2}
displays the obtained measures.
For comparison,
result values corresponding to the following
transformations were also included:
the 8-point exact DCT,
the 8-point transforms employed in AVC/H.264~\cite{H264_8transform} and HEVC/H.265~\cite{hevc1,sullivan2012},
the 8-point SDCT~\cite{haweel2001new},
and
the approximation described in~\cite{multibeam2012}.
Although these additional transforms are in the Feig-Winograd subspace,
they are not a result of the proposed optimization problem
detailed in Section~\ref{S:results}.

\begin{table}
\centering
\caption{Assessment of Feig-Winograd matrices}
\label{t:cases2}
\begin{tabular}{cccccccc}
\toprule
Transform\!\!\!\! & 
$\epsilon(\bm{\alpha})$ \!\!\!\!\!\! & 
\!\!$\operatorname{MSE}(\bm{\alpha})$ \!\!\!\!\!\!\!\!&
$C_g(\bm{\alpha})$ \!\!\!\!\!\!& 
$\eta(\bm{\alpha})$ & 
\!\!\!\! $\mathcal{A}(\bm{\alpha})$ \!\!\!\! & \!\!\!\! $\mathcal{S}(\bm{\alpha})$ \!\!\!\!& Mult.\\
\midrule
\multicolumn{8}{c}{Optimal transforms} \\
\midrule
$\bm{\mathsf{T}}^{(1)}_8$~\cite{lengwehasatit2004scalable}& 0.870 & 0.006 & 8.39 & 88.70& 24 & 2&0\\
$\bm{\mathsf{T}}^{(2)}_8$~\cite{cb2011}& 1.794 & 0.010 & 8.18 & 87.43& 22 & 0&0\\
$\bm{\mathsf{T}}^{(3)}_8$~\cite{bc2012} & 8.659 & 0.059 & 7.33 & 80.90& 14 & 0&0\\
$\bm{\mathsf{T}}^{(4)}_8$ & 7.734 & 0.056 & 7.54 & 81.99& 16 & 2&0\\
$\bm{\mathsf{T}}^{(5)}_8$ & 8.659 & 0.059 & 7.37 & 81.18 & 18 & 0&0 \\
$\bm{\mathsf{T}}^{(6)}_8$ & 7.734 & 0.055 & 7.58 & 82.27 & 20 & 2&0 \\
$\bm{\mathsf{T}}^{(7)}_8$ & 7.532 & 0.054 & 7.56 & 82.70 & 20 & 6&0 \\
$\bm{\mathsf{T}}^{(8)}_8$ & 7.414 & 0.053 & 7.58 & 83.08 & 20 & 10&0 \\
$\bm{\mathsf{T}}^{(16)}_8$ & 3.316 & 0.021 & 6.05 & 83.08& 18 & 0&0 \\
\midrule
\multicolumn{8}{c}{Non-optimal transforms} \\
\midrule
SDCT~\cite{haweel2001new}&3.316&0.021&6.03&82.62&24&0&0\\
Approximation in~\cite{multibeam2012}&0.870&0.006&8.34&88.06&24&6&0\\
AVC~\cite{H264_8transform}
&0.072&0.000&8.78&92.46&32&10&0\\
HEVC~\cite{Fuldseth2011,Meher2014} &0.002&0.000&8.82&93.82&28&0&22\\
DCT&0&0&8.83&93.99&28&0&22\\
\bottomrule
\end{tabular} 
\end{table}

\subsection{Discussion and Comparison}

Measurements shown in Table~\ref{t:cases2} 
revealed
that, 
among the optimal Feig-Winograd matrices, 
the transform $\bm{\mathsf{T}}^{(1)}_8$ 
presents superior performance
according to all proximity measures. 
We recall that $\bm{\mathsf{T}}^{(1)}_8$
coincides with DCT approximation
proposed
by Lengwehasatit-Ortega
which is well-known 
for being a very good approximation~\cite{lengwehasatit2004scalable}.
Nevertheless,
it is also recognized for
its comparatively high computational complexity,
which makes it less attractive a tool
when other approximations are considered.
Approximation~$\bm{\mathsf{T}}^{(2)}_8$,
identified as the rounded DCT~\cite{cb2011},
also
displays close mathematical proximity to the DCT,
while demanding
lower computational effort in comparison to~$\bm{\mathsf{T}}^{(1)}_8$.
Previously
described in~\cite{bc2012},
matrix~$\bm{\mathsf{T}}^{(3)}_8$
has the distinction of
requiring only 14~additions and no bit-shifting operation,
achieving the minimal possible arithmetic complexity 
among all considered Feig-Winograd matrices 
(cf.~\eqref{eq:additions}, \eqref{eq:bit-shifting}).

Regarding the 
new proposed approximations, 
matrix $\bm{\mathsf{T}}^{(4)}_8$ 
possesses 
good performance in terms of proximity measures,
while requiring only 16~additions.
New matrices
$\bm{\mathsf{T}}^{(5)}_8$ and $\bm{\mathsf{T}}^{(16)}_8$
require only 18 additions
and can be more closely compared.
While $\bm{\mathsf{T}}^{(5)}_8$ leads to an orthogonal approximation,
$\bm{\mathsf{T}}^{(16)}_8$ furnishes an nonorthogonal one.
Matrix~$\bm{\mathsf{T}}^{(16)}_8$ 
has the smallest total error energy 
and 
mean square error among all new transformations.
In terms of unified coding gain,
the orthogonal transform $\bm{\mathsf{T}}^{(5)}_8$ 
outperforms $\bm{\mathsf{T}}^{(16)}_8$;
on the other hand,
when transform efficiency is considered,
the nonorthogonal approximation excels.

New efficient 
transforms~$\bm{\mathsf{T}}^{(6)}_8$, $\bm{\mathsf{T}}^{(7)}_8$, and $\bm{\mathsf{T}}^{(8)}_8$
exhibit good coding-related figures,
whereas
they impose higher computational complexity requirements.
At the same time,
their
additive complexity is not as large as the one required by 
the
Lengwehasatit-Ortega ($\bm{\mathsf{T}}^{(1)}_8$)
approximation.

Additionally,
Table~\ref{t:cases2} 
informs
that a trade-off
between
computational cost and performance is observed in some rough sense.
To further our analysis,
we also examined
the following 
low-complexity transformations
which are not encompassed
in the Feig-Winograd matrix space:
the Walsh-Hadamard transform (WHT)~\cite{Horadam2007}
and 
the series of DCT approximations 
introduced by Bouguezel-Ahmad-Swamy
labeled as
$\mbox{BAS}^{(1)}$~\cite{Bouguezel2008}, 
$\mbox{BAS}^{(2)}$~\cite{bas2008}, 
$\mbox{BAS}^{(3)}$~\cite{bas2009}, 
$\mbox{BAS}^{(4)}$~\cite{bas2010}, 
$\mbox{BAS}^{(5)}$~\cite{bas2011} (for $a=1$), 
$\mbox{BAS}^{(6)}$~\cite{bas2011} (for $a=0$), 
$\mbox{BAS}^{(7)}$~\cite{bas2011} (for $a=0.5$),
and 
$\mbox{BAS}^{(8)}$~\cite{bas2013}.
Table~\ref{t:nao_FW}
lists
the assessment measurements
for the above-mentioned approximations.
The new proposed matrix~$\bm{\mathsf{T}}^{(16)}_8$ 
outperforms the BAS series approximations
in terms of total error energy and MSE.

\begin{table}
 \centering
 \caption{Assessment of selected non-Feig-Winograd Transforms}
 \label{t:nao_FW}
 \begin{tabular}{cccccccc}
 \toprule
  \!\!\!\!\!\scriptsize{Transform}\!\!\!&\!\!\!\!\scriptsize{Orthogonalizable?}\!\!\!\!\!&\!\!\!\!\!$\epsilon(\bm{\alpha})$ \!\!\!\!\!\! & 
\!\!\!\!\!$\operatorname{MSE}(\bm{\alpha})$ \!\!\!\!\!&
\!\!\!\!$C_g(\bm{\alpha})$ \!\!\!\!\!\!& 
\!\!\!\!\!$\eta(\bm{\alpha})$\!\!\!\!\!& 
\!\!\!\!\!\! $\mathcal{A}(\bm{\alpha})$\!\!\!\! &\!\!\!$\mathcal{S}(\bm{\alpha})$\!\!\!\!\! \\
  \midrule
  $\mbox{BAS}^{(1)}$&No&4.188&0.019&6.27&83.17&21&0\\
  $\mbox{BAS}^{(2)}$&Yes&5.929&0.024&8.12&86.86&18&2\\
  $\mbox{BAS}^{(3)}$&Yes&6.854&0.028&7.91&85.38&18&0\\
  $\mbox{BAS}^{(4)}$&Yes&4.093&0.021&8.33&88.22&24&4\\
  $\mbox{BAS}^{(5)}$&Yes&26.864&0.071&7.91&85.38&18&0\\
  $\mbox{BAS}^{(6)}$&Yes&26.864&0.071&7.91&85.64&16&0\\
  $\mbox{BAS}^{(7)}$&Yes&26.402&0.068&8.12&86.86&18&2\\
  $\mbox{BAS}^{(8)}$&Yes&35.064&0.102&7.95&85.31&24&0\\
  WHT&Yes&5.049&0.025&7.95&85.31&24&0\\
  \bottomrule
 \end{tabular}
\end{table}

\begin{figure*}%
\centering
\subfigure[$\epsilon(\bm{\alpha})\,\times \,\mathcal{A}(\bm{\alpha})$]
{\includegraphics[width=0.48\linewidth]{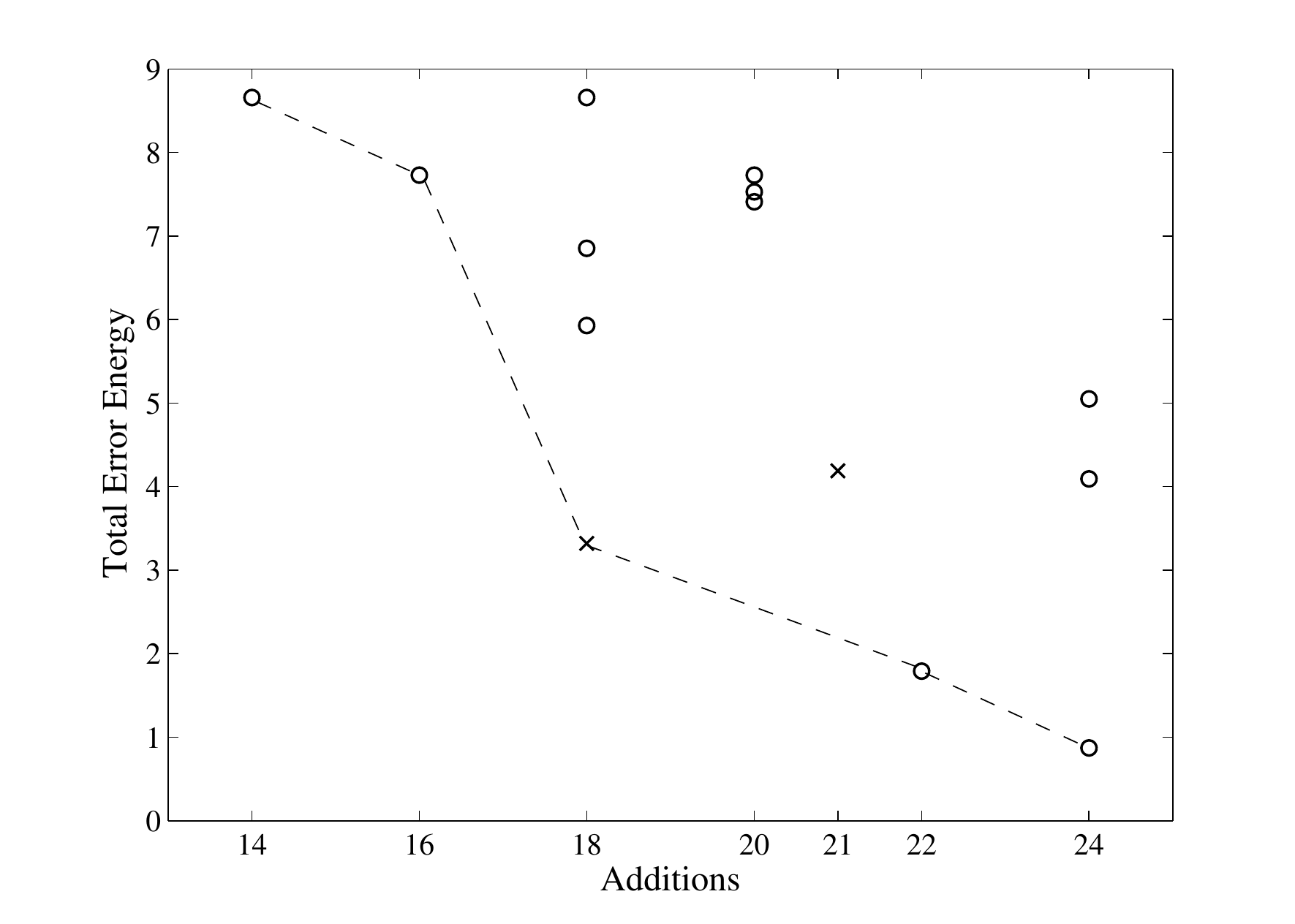} 
\put(-188,133){\scriptsize $\bm{\mathsf{T}}^{(3)}_8$}
\put(-127,136){\scriptsize $\bm{\mathsf{T}}^{(5)}_8$}
\put(-109,125){\scriptsize $\bm{\mathsf{T}}^{(7)}_8$}
\put(-96,117){\scriptsize $\bm{\mathsf{T}}^{(8)}_8$}
\put(-96,133){\scriptsize $\bm{\mathsf{T}}^{(6)}_8$}
\put(-164,121){\scriptsize $\bm{\mathsf{T}}^{(4)}_8$}
\put(-133,55){\scriptsize $\bm{\mathsf{T}}^{(16)}_8$}
\put(-72,33){\scriptsize $\bm{\mathsf{T}}^{(2)}_8$}
\put(-43,21){\scriptsize $\bm{\mathsf{T}}^{(1)}_8$}
\put(-45,95){\scriptsize WHT}
\put(-47,70){\scriptsize $\mbox{BAS}^{(4)}$}
\put(-89,71){\scriptsize $\mbox{BAS}^{(1)}$}
\put(-134,95){\scriptsize $\mbox{BAS}^{(2)}$}
\put(-136,120){\scriptsize $\mbox{BAS}^{(3)}$}
\label{fw_error}}
\subfigure[$\operatorname{MSE}(\bm{\alpha})\,\times \,\mathcal{A}(\bm{\alpha})$]
{\includegraphics[width=0.48\linewidth]{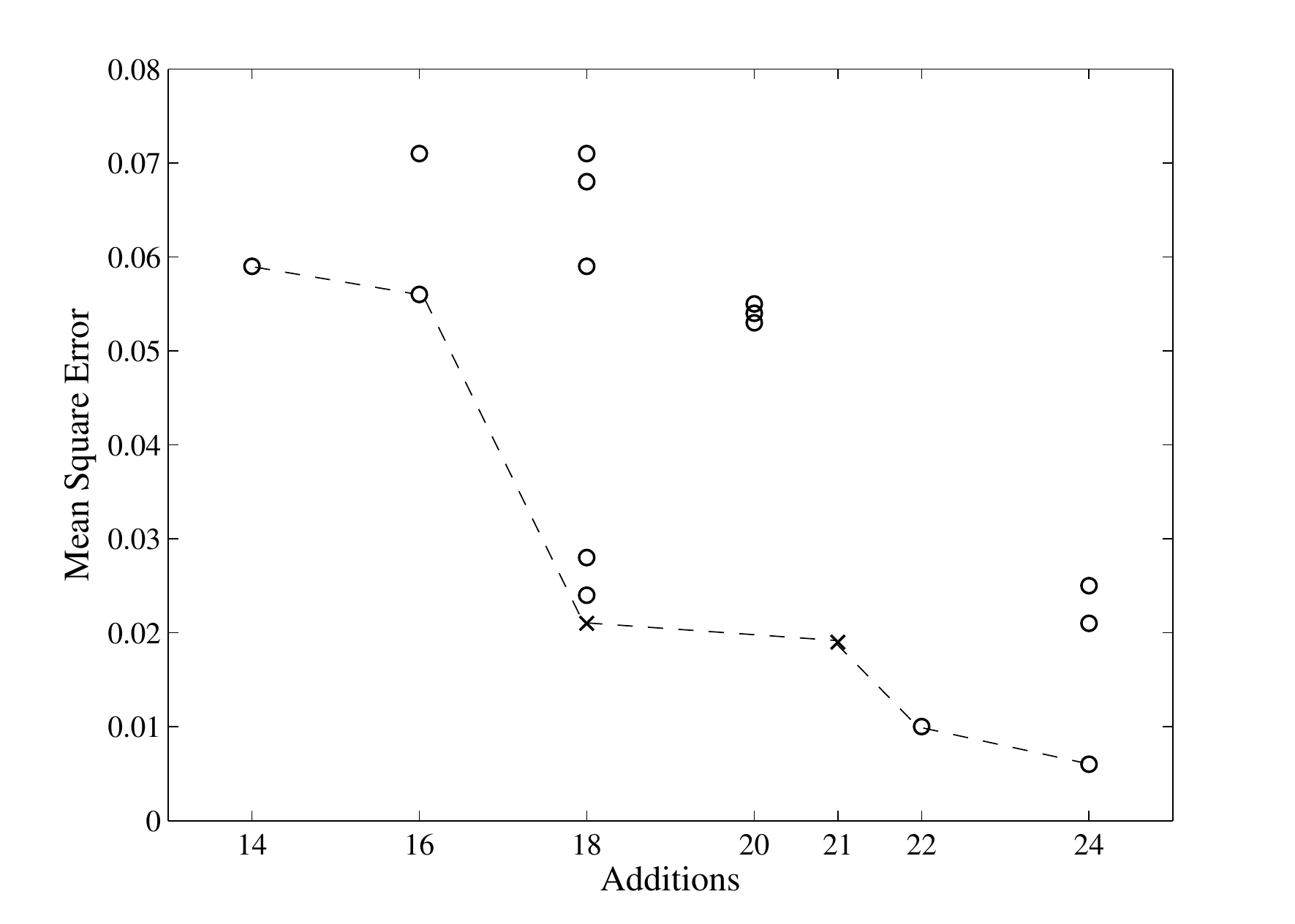} 
\put(-190,105){\scriptsize $\bm{\mathsf{T}}^{(3)}_8$}
\put(-128,106){\scriptsize $\bm{\mathsf{T}}^{(5)}_8$}
\put(-109,105){\scriptsize $\bm{\mathsf{T}}^{(7)}_8$}
\put(-98,97){\scriptsize $\bm{\mathsf{T}}^{(8)}_8$}
\put(-98,112){\scriptsize $\bm{\mathsf{T}}^{(6)}_8$}
\put(-164,100){\scriptsize $\bm{\mathsf{T}}^{(4)}_8$}
\put(-132,43){\scriptsize $\bm{\mathsf{T}}^{(16)}_8$}
\put(-73,24){\scriptsize $\bm{\mathsf{T}}^{(2)}_8$}
\put(-50,21){\scriptsize $\bm{\mathsf{T}}^{(1)}_8$}
\put(-47,62){\scriptsize WHT}
\put(-47,44){\scriptsize $\mbox{BAS}^{(4)}$}
\put(-100,40){\scriptsize $\mbox{BAS}^{(1)}$}
\put(-120,55){\scriptsize $\mbox{BAS}^{(2)}$}
\put(-129,67){\scriptsize $\mbox{BAS}^{(3)}$}
\put(-149,124){\scriptsize $\mbox{BAS}^{(7)}$}
\put(-133,138){\scriptsize $\mbox{BAS}^{(5)}$}
\put(-163,137){\scriptsize $\mbox{BAS}^{(6)}$}
\label{fw_mse}}\\
\subfigure[$C_g(\bm{\alpha})\,\times \,\mathcal{A}(\bm{\alpha})$]
{\includegraphics[width=0.48\linewidth]{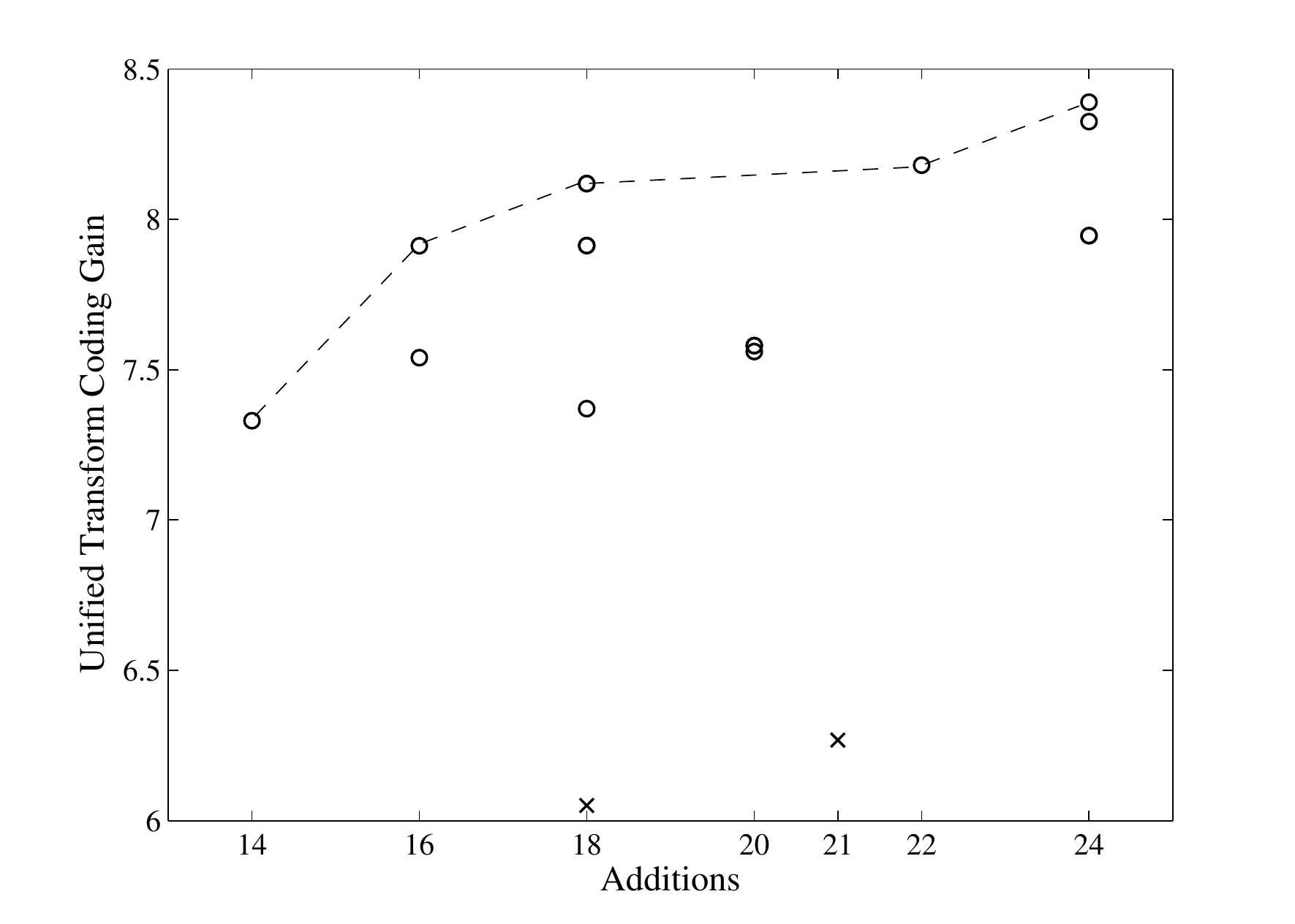} 
\put(-191,78){\scriptsize $\bm{\mathsf{T}}^{(3)}_8$}
\put(-158,90){\scriptsize $\bm{\mathsf{T}}^{(4)}_8$}
\put(-133,26){\scriptsize $\bm{\mathsf{T}}^{(16)}_8$}
\put(-130,80){\scriptsize $\bm{\mathsf{T}}^{(5)}_8$}
\put(-105,104){\scriptsize $\bm{\mathsf{T}}^{(6)}_8,\bm{\mathsf{T}}^{(8)}_8$}
\put(-98,91){\scriptsize $\bm{\mathsf{T}}^{(7)}_8$}
\put(-73,136){\scriptsize $\bm{\mathsf{T}}^{(2)}_8$}
\put(-35,141){\scriptsize $\bm{\mathsf{T}}^{(1)}_8$}
\put(-70,112){\scriptsize WHT, $\mbox{BAS}^{(8)}$}
\put(-46,129){\scriptsize $\mbox{BAS}^{(4)}$}
\put(-93,36){\scriptsize $\mbox{BAS}^{(1)}$}
\put(-143,133){\scriptsize $\mbox{BAS}^{(2)}, \mbox{BAS}^{(7)}$}
\put(-146,110){\scriptsize $\mbox{BAS}^{(3)}, \mbox{BAS}^{(5)}$}
\put(-174,118){\scriptsize $\mbox{BAS}^{(6)}$}
\label{fw_Cg}}
\subfigure[$\eta(\bm{\alpha})\,\times \,\mathcal{A}(\bm{\alpha})$]
{\includegraphics[width=0.48\linewidth]{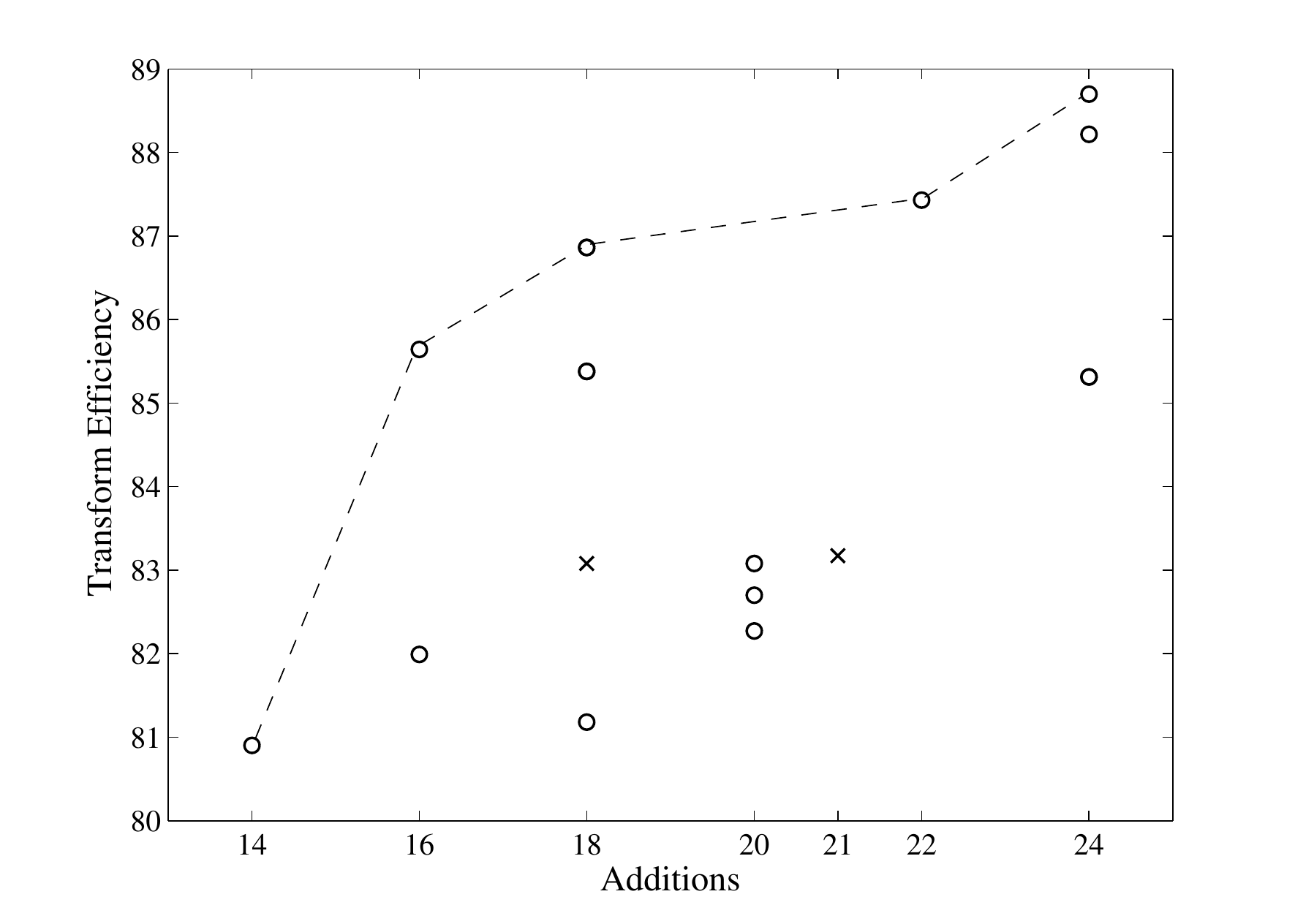} 
\put(-189,22){\scriptsize $\bm{\mathsf{T}}^{(3)}_8$}
\put(-161,37){\scriptsize $\bm{\mathsf{T}}^{(4)}_8$}
\put(-132,53){\scriptsize $\bm{\mathsf{T}}^{(16)}_8$}
\put(-130,25){\scriptsize $\bm{\mathsf{T}}^{(5)}_8$}
\put(-100,67){\scriptsize $\bm{\mathsf{T}}^{(8)}_8$}
\put(-110,55){\scriptsize $\bm{\mathsf{T}}^{(7)}_8$}
\put(-98,42){\scriptsize $\bm{\mathsf{T}}^{(6)}_8$}
\put(-74,130){\scriptsize $\bm{\mathsf{T}}^{(2)}_8$}
\put(-34,140){\scriptsize $\bm{\mathsf{T}}^{(1)}_8$}
\put(-67,87){\scriptsize WHT, $\mbox{BAS}^{(8)}$}
\put(-45,127){\scriptsize $\mbox{BAS}^{(4)}$}
\put(-86,56){\scriptsize $\mbox{BAS}^{(1)}$}
\put(-159,122){\scriptsize $\mbox{BAS}^{(2)}, \mbox{BAS}^{(7)}$}
\put(-145,88){\scriptsize $\mbox{BAS}^{(3)}, \mbox{BAS}^{(5)}$}
\put(-174,102){\scriptsize $\mbox{BAS}^{(6)}$}
\label{fw_eff}}
\caption{
Assessment plots for
the proposed efficient approximations;
competing methods were included for comparison.}
\label{f:quality}
\end{figure*}
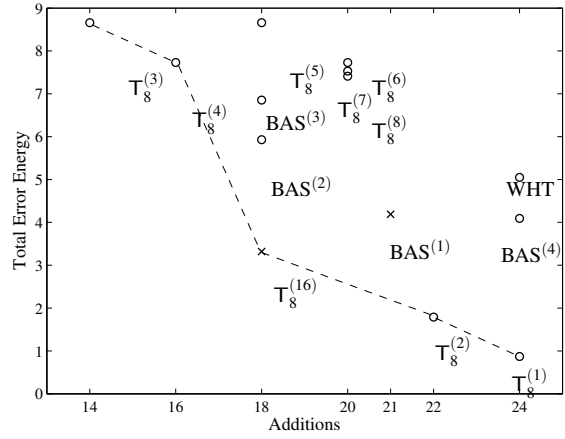
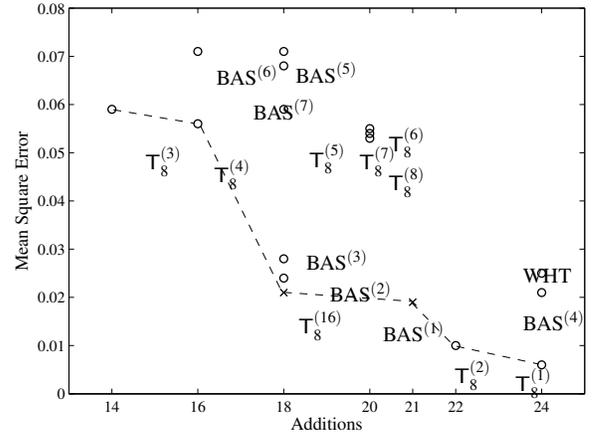
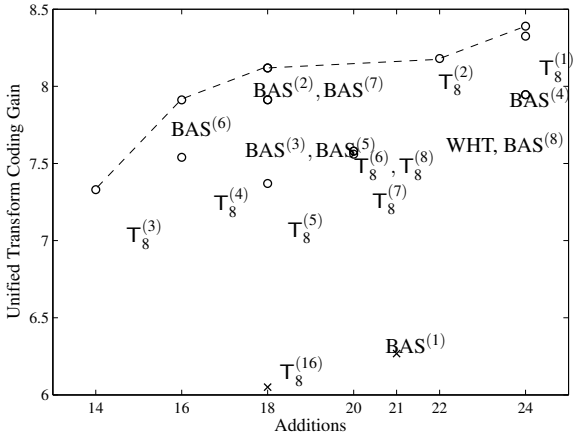
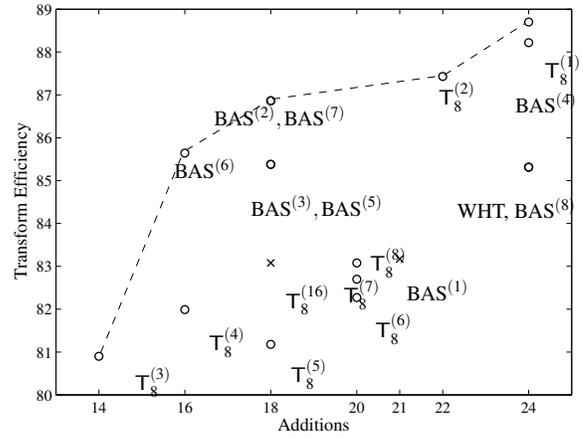

To better visualize such balance,
we devised 
scatter plots
relating to
computational cost
with
the discussed performance measures.
Fig.~\ref{f:quality} displays the resulting scatter
plots
where 
each transform corresponds
to a labeled point.
Orthogonal transforms are
denoted by circle marks
and
nonorthogonal transforms
are indicated by cross marks.
Approximations
$\mbox{BAS}^{(5)}$, 
$\mbox{BAS}^{(6)}$, 
$\mbox{BAS}^{(7)}$, 
and 
$\mbox{BAS}^{(8)}$
were not included
in
Fig.~\ref{f:quality}(a)
because
their
measurements
were exceedingly large.
Similarly,
$\mbox{BAS}^{(8)}$ transform 
was
also excluded from Fig.~\ref{f:quality}(b).

Fig.~\ref{f:quality}(a)--(b)
indicates that
the Feig-Winograd approximations
excel in terms of proximity to
the original DCT transforms
as measured 
by the
total error energy and MSE. 
Additionally,
the Feig-Winograd transforms
presented better coding-related figures,
except 
for 
transformations that
require
16 and 18 additions.

In particular,
considering
transformations that required only
16 additions,
we notice that 
the proposed approximation
$\bm{\mathsf{T}}^{(4)}_8$ 
outperformed
$\mbox{BAS}^{(6)}$ 
in terms of
total error energy and MSE.
However, 
$\mbox{BAS}^{(6)}$ shows better coding gain performance
than $\bm{\mathsf{T}}^{(4)}_8$.
Now
considering
transformations demanding
18 additions,
we have that
approximations 
$\mbox{BAS}^{(2)}$, 
$\mbox{BAS}^{(3)}$, 
$\mbox{BAS}^{(5)}$,
and 
$\mbox{BAS}^{(7)}$ 
showed better coding gain performance
than the proposed
transformations $\bm{\mathsf{T}}^{(5)}_8$ and $\bm{\mathsf{T}}^{(6)}_8$.
As expected,
the approximation
$\bm{\mathsf{T}}^{(1)}_8$ 
has 
the best coding performance 
among all considered transforms in Fig.~\ref{f:quality}.
Suggested approximation $\bm{\mathsf{T}}^{(2)}_8$
could
outperform
non-Feig-Winograd transform
in all metrics.

In terms of the nonorthogonal transforms
$\bm{\mathsf{T}}^{(16)}_8$ and $\mbox{BAS}^{(1)}$,
we report that
$\bm{\mathsf{T}}^{(16)}_8$ possesses
the smaller total error energy
when compared to $\mbox{BAS}^{(1)}$,
while
presenting
a similar performance. 
However,
$\bm{\mathsf{T}}^{(16)}_8$
is 14.3\% less computationally complex.
Also, as discussed in~\cite{Katto1991}, 
it is generally expected that
nonorthogonal transforms 
present lower values of coding gain compared
with orthogonal transforms (cf. Fig.~\ref{f:quality}(c)). 
This is expected
since coding gain measures are optimized for the KLT~\cite{Goyal2001}, 
which is orthogonal.

We also notice
that
Figs.~\ref{fw_error}--\subref{fw_eff} 
can be 
interpreted as
results of
four different 
bi-objective optimization problems~\cite[p.~245]{Miettinen1999},
where
the 
trade-off between computational cost and performance measures are 
emphasized. 
The optimal solutions of
the bi-objective optimization problems 
are situated on the boundary 
of
set of feasible solutions~\cite[p.~28]{Ehrgott2000}. 
Such boundaries
are shown in Fig.~\ref{f:quality}
and 
roughly represent
the \emph{Pareto} frontiers for each problem~\cite[p.~11]{Miettinen1999}.
Thus, in this sense, 
we separate the transforms situated on each Pareto frontier 
as
\emph{optimal} transformations.
Such transforms were: 
$\bm{\mathsf{T}}^{(1)}_8$, 
$\bm{\mathsf{T}}^{(2)}_8$, 
$\bm{\mathsf{T}}^{(3)}_8$, 
$\bm{\mathsf{T}}^{(4)}_8$, 
$\bm{\mathsf{T}}^{(16)}_8$, 
$\mbox{BAS}^{(1)}$, 
$\mbox{BAS}^{(2)}$, 
$\mbox{BAS}^{(6)}$,
and 
$\mbox{BAS}^{(7)}$.
This reduced set of transforms
was submitted 
to subsequent image compression analysis.

However,
notice that although
the remaining
transforms 
$\bm{\mathsf{T}}^{(5)}_8$, $\bm{\mathsf{T}}^{(6)}_8$, \ldots, $\bm{\mathsf{T}}^{(15)}_8$ 
were outperformed,
the performance of a DCT approximation
is highly dependent on the envisioned task. 
Thus, 
these transforms may be adequate in other contexts, 
as suggested in~\cite{bas2013}.

In terms of comparisons
with other methods,
we notice that
current video standards 
AVC/H.264~\cite{wiegand2003, H264_8transform}
and 
HEVC/H.265~\cite{hevc1, Fuldseth2011, Meher2014} 
employ integer DCT approximations,
which are indeed encompassed in the Feig-Winograd matrix subspace. 
However, 
these integer matrices 
do not satisfy the low-complexity restriction on 
the matrix elements---namely
having its parameter vector defined over the set $\mathcal{P}=\{0, \pm1/2, \pm1, \pm2 \}$,
as shown in Section~\ref{cases}.
Possessing large elements,
the implementation of such matrices requires
more sophisticate operation %
schemes
which make them more computationally expensive~\cite{H264_8transform, Fuldseth2011, Meher2014}
when compared to the extremely low-complexity approximations discussed here.
For instance,
in terms of additive complexity,
the AVC requires from 50.00\% to 128.57\% more operations
in comparison with the optimal transformations, 
as showed in Table~\ref{t:cases2}.
On its turn, 
HEVC requires multiplication operations,
which are much more expensive than simple additions~\cite{Blahut2010}.
For a fair comparison, 
we restricted our subsequent analyses to 
the set of very low-complexity matrices
whose elements are in~$\mathcal{P}$.  

\section{Image Compression}
\label{S:image-compression}

In order to evaluate the performance 
of the selected approximations
in image compression,
we adopted
a JPEG-like computational simulation~\cite{Wallace1992,penn1992,
bas2008, Bouguezel2008, bas2009, bas2010, bas2011, bas2013, 
cb2011,
bc2012, 
multibeam2012, 
bayer201216pt, 
Edirisuriya2012,
Potluri2013}
based on
a set of 45 512$\times$512 8-bit greyscale standard images
obtained from a public image bank~\cite{uscsipi}.
Images were split
into 8$\times$8 sub-blocks,
which were submitted to a given \mbox{2-D}
transformation~\cite{Wallace1992,suzuki2010integer}
depending on the considered DCT approximation.
The \mbox{2-D}
transform operation depends
on
whether the considered transform is
orthogonal or not.
Let $\mathbf{A}$ be a 8$\times$8 sub-block of the considered image.
In general,
the \mbox{2-D} approximate DCT of~$\mathbf{A}$
be written
as~\cite{suzuki2010integer}:
\begin{align*}
\mathbf{B} 
&=
\begin{cases}
\hat{\mathbf{C}}_8 \cdot \mathbf{A} \cdot \hat{\mathbf{C}}_8^{\top},
&
\text{if $\mathbf{T}^{(\bm{\alpha})}_8$ satisfies~\eqref{equation-condition-1},}
\\
\hat{\mathbf{C}}_8 \cdot \mathbf{A} \cdot \hat{\mathbf{C}}_8^{-1},
&
\text{otherwise.}
\end{cases}
\end{align*}
This computation furnished 64 coefficients 
in the approximate transform domain
for each sub-block.
According to the standard zigzag sequence~\cite{Wallace1992},
only the $r$ initial coefficients in each block 
were retained 
and
employed to reconstruct the image~\cite{cb2011}. 
All the remaining coefficients were set to zero.
We adopted $1 \leq r \leq 45$. 
Based on 8-bit images, 
this approach implies in the fixed bitrate equals to $r/8$~bpp.
After compression,
the inverse \mbox{2-D}
transform
was applied to reconstruct the processed data.
Subsequently,
image quality was evaluated.

Above outlined methodology is also
described in~\cite{haweel2001new}
and 
supported 
in~\cite{bas2008, Bouguezel2008, bas2009, bas2010, bas2011}.
However,
in contrast to the 
JPEG-based 
image compression experiments described in~\cite{haweel2001new,bas2008, Bouguezel2008, bas2009, bas2010, bas2011}, 
we
adopted 
average image quality measures
from 
the entire image set~\cite{cb2011,bc2012,bayer201216pt}
instead of the results obtained from particular selected images. 
Thus, 
our approach is less prone 
to variance effects and fortuitous input data, 
being more robust a methodology~\cite{kay1993}.

Image degradation was assessed 
by means of 
the peak signal-to-noise ratio (PSNR)~\cite{Huynh-Thu2008} 
and 
the structural similarity index~(SSIM)~\cite{Wang2004}.
The PSNR is a quality measure widely used 
in 
image processing~\cite{Huynh-Thu2008}
and
the SSIM is regarded
as a complementary
method for image quality assessment~\cite{Wang2004}.
In fact,
the SSIM considers 
luminance, contrast, and image structure
to quantify image degradation,
being consistent with 
subjective quality measurements~\cite{Wang2011}.

\begin{figure*}%
 \centering
 \subfigure[]
 {\includegraphics[width=0.475\linewidth]{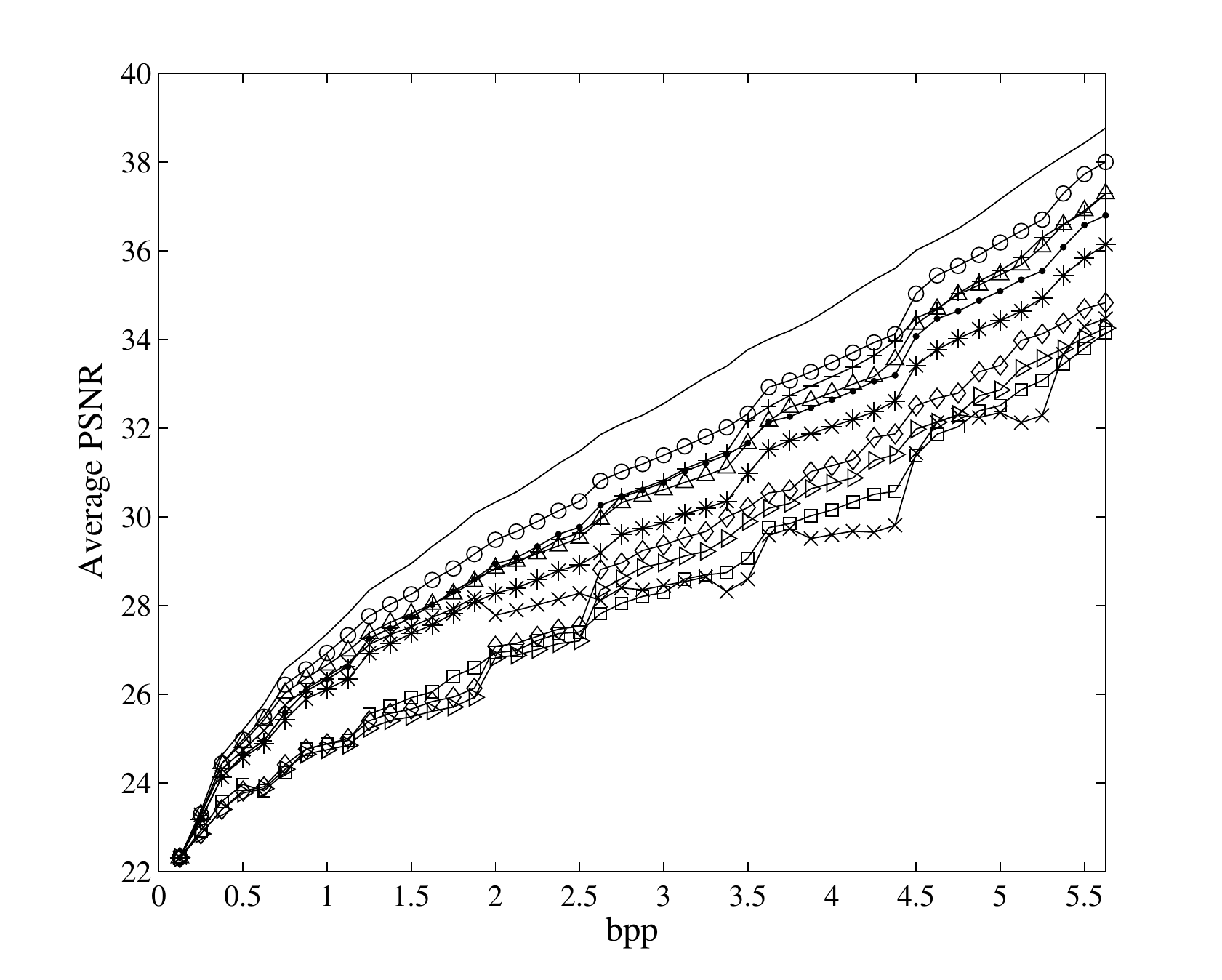}\label{psnr3}}
 \subfigure[]
 {\includegraphics[width=0.475\linewidth]{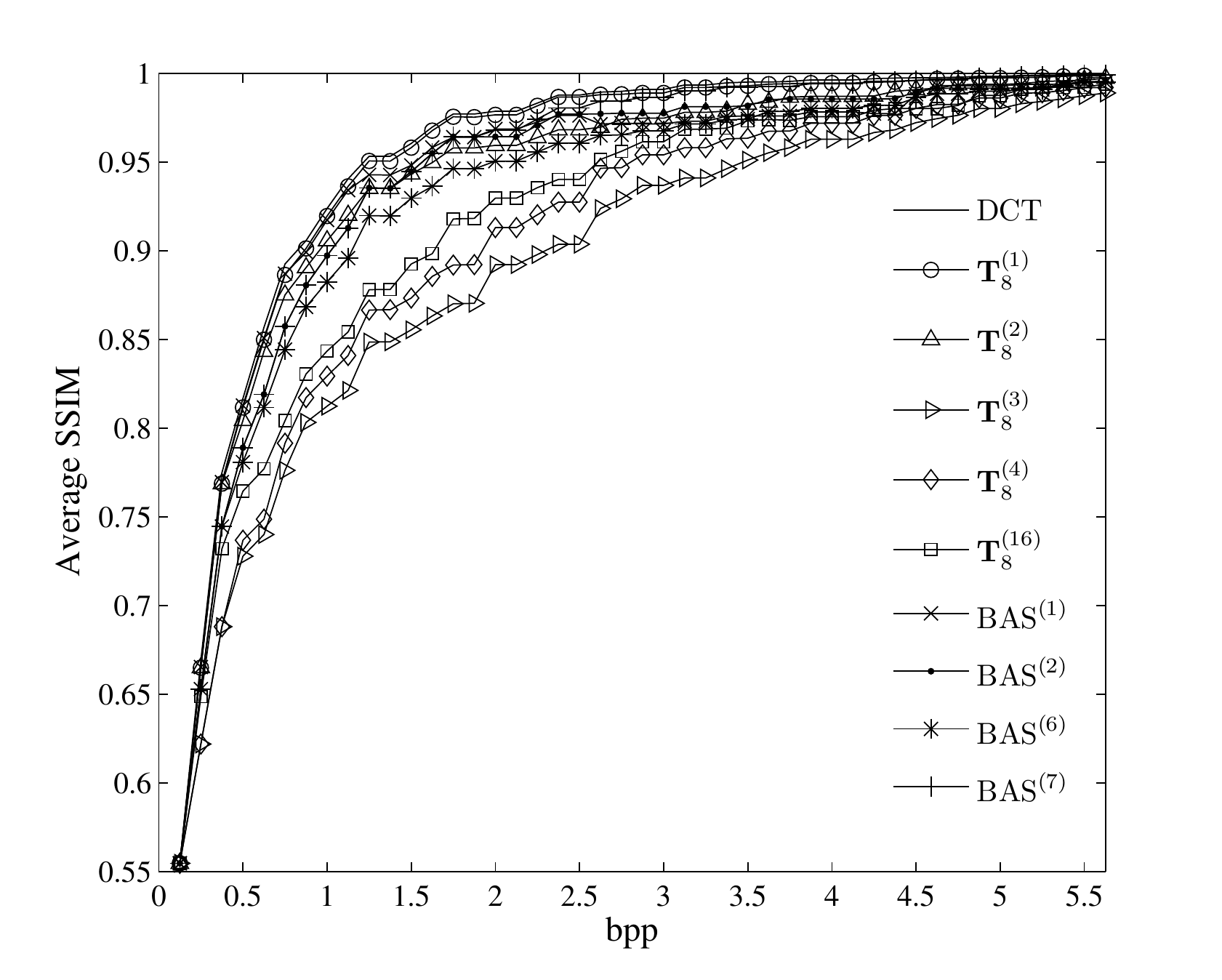}\label{ssim3}} 
 \subfigure[]
 {\includegraphics[width=0.475\linewidth]{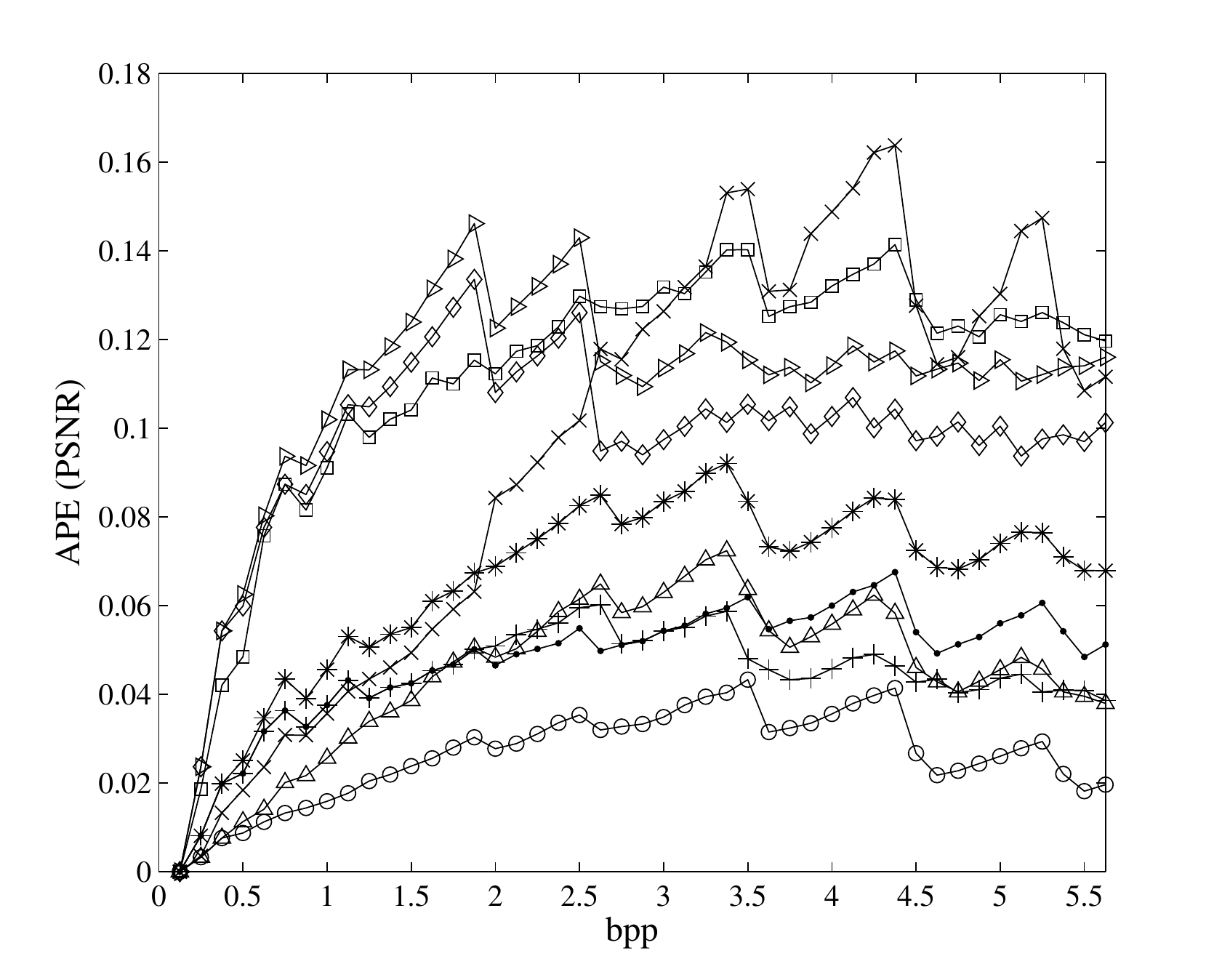}\label{psnr_ape3}}
 \subfigure[]
 {\includegraphics[width=0.475\linewidth]{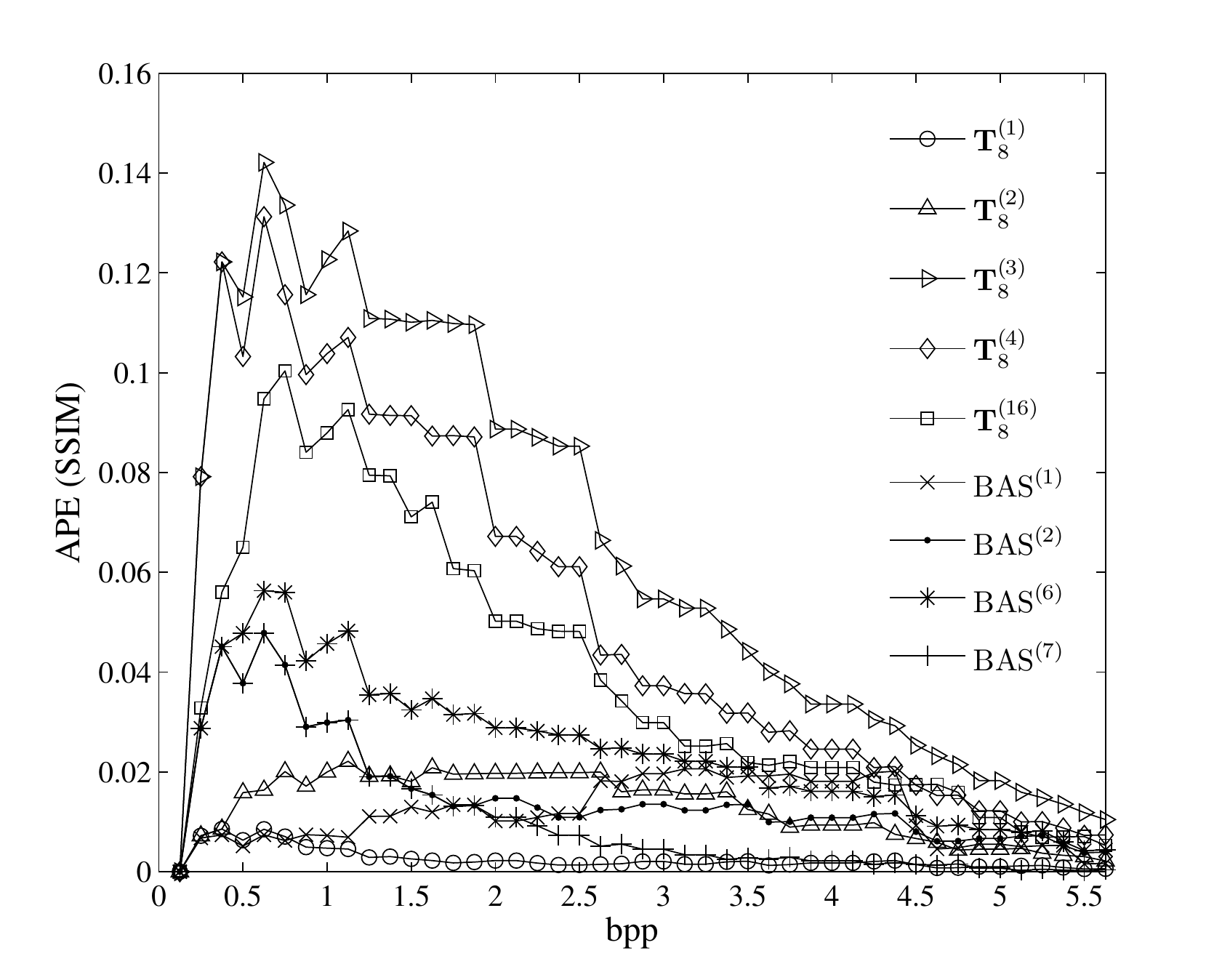}\label{ssim_ape3}}
 \caption{
Quality measures of selected optimal approximations
for several values of bpp according to the following
figures of merit:
\subref{psnr3}~Average PSNR, 
\subref{psnr_ape3}~Average PSNR absolute percentage error relative to the DCT, 
\subref{ssim3}~Average SSIM, 
and 
\subref{ssim_ape3}~Average SSIM absolute percentage error relative to the DCT.}
 \label{f:measures2}
\end{figure*}

Figs.~\ref{f:measures2}(a)--(b) show 
the obtained quality measures 
for
the transforms identified as optimal
according to the discussion 
at the end of
the preceding section.
This type of performance curves
is commonly employed
as a comparison tool
in approximate DCT 
literature~\cite{Bouguezel2008,bas2009,bas2011,cb2011,bc2012}.
For an improved visualization
of the performance curves,
we considered
the
absolute percentage error (APE) relative to the DCT of 
both the PSNR and SSIM measurements.
The resulting values 
are plotted
in
Figs.~\ref{f:measures2}(c)--(d).
To further validate
our methodology,
we also computed additional descriptive statistical measures
of the results.
In particular,
the coefficient of variation~\cite{Abdi2010}
was found to be consistently less than~$16\%$,
for all data sample.
This fact suggests that adopting
average values for the metrics
is indeed an adequate approach,
as supported in~\cite[p.~155]{Brown1998} and~\cite[p.~387]{Wackerly2007}.

As expected,
among all considered methods, 
$\bm{\mathsf{T}}^{(1)}_8$ 
possesses the best performance in terms of image quality
at the cost of the highest computational 
complexity~\cite{Bouguezel2008,bas2009,bas2010}.
Feig-Winograd approximation $\bm{\mathsf{T}}^{(2)}_8$ 
could outperform 
transformations
$\bm{\mathsf{T}}^{(3)}_8$, 
$\bm{\mathsf{T}}^{(4)}_8$, 
$\bm{\mathsf{T}}^{(16)}_8$,
and 
$\mbox{BAS}^{(6)}$, 
while exhibiting
similar performance to
$\mbox{BAS}^{(2)}$ and $\mbox{BAS}^{(7)}$.
Transformations~$\bm{\mathsf{T}}^{(3)}_8$,
$\bm{\mathsf{T}}^{(4)}_8$,
and
$\bm{\mathsf{T}}^{(16)}_8$
also 
performed better than
$\mbox{BAS}^{(1)}$
in
terms of PSNR,
for $r>25$ 
(bitrate larger than 3.125~bpp),
and of SSIM,
for all values of $r$.
The new nonorthogonal $\bm{\mathsf{T}}^{(16)}_8$ 
outperformed $\bm{\mathsf{T}}^{(3)}_8$ and $\bm{\mathsf{T}}^{(4)}_8$.
However,
transformation $\bm{\mathsf{T}}^{(16)}_8$ 
showed a lower arithmetic complexity 
when compared to
the also nonorthogonal
$\mbox{BAS}^{(1)}$.
Proposed
approximation
$\bm{\mathsf{T}}^{(4)}_8$ 
surpassed
$\bm{\mathsf{T}}^{(3)}_8$ 
in both PSNR and SSIM measurements
for
all compression ratio,
requiring only two extra additions for its computation. 

\begin{figure*}
\centering
\subfigure[$\bm{\mathsf{T}}^{(1)}_8$ (PSNR=35.176, SSIM=0.995)]
{\includegraphics[width=0.18\linewidth]{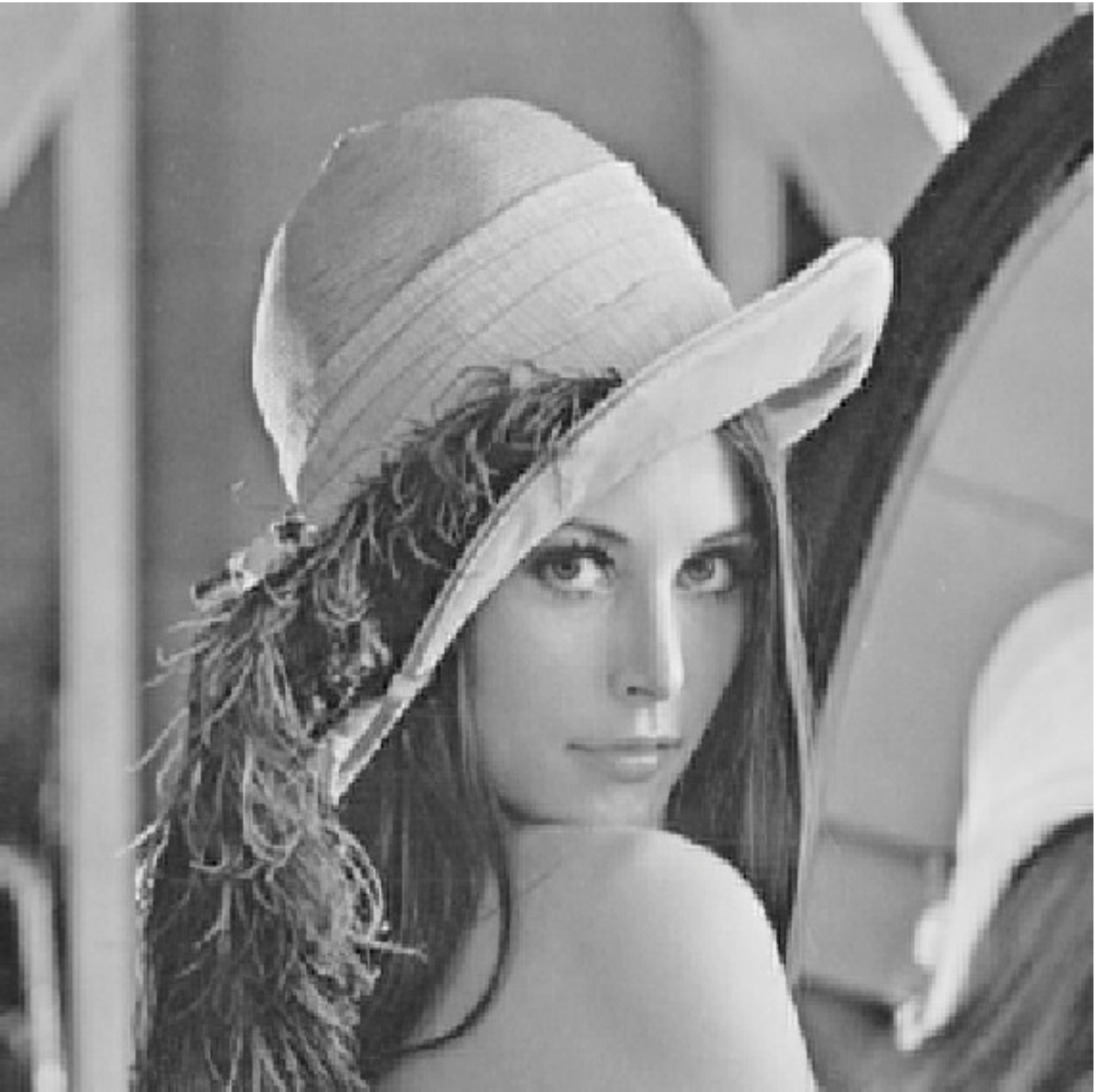} \label{lena_t1}}
\subfigure[$\bm{\mathsf{T}}^{(2)}_8$ (PSNR=34.138, SSIM=0.989)]
{\includegraphics[width=0.18\linewidth]{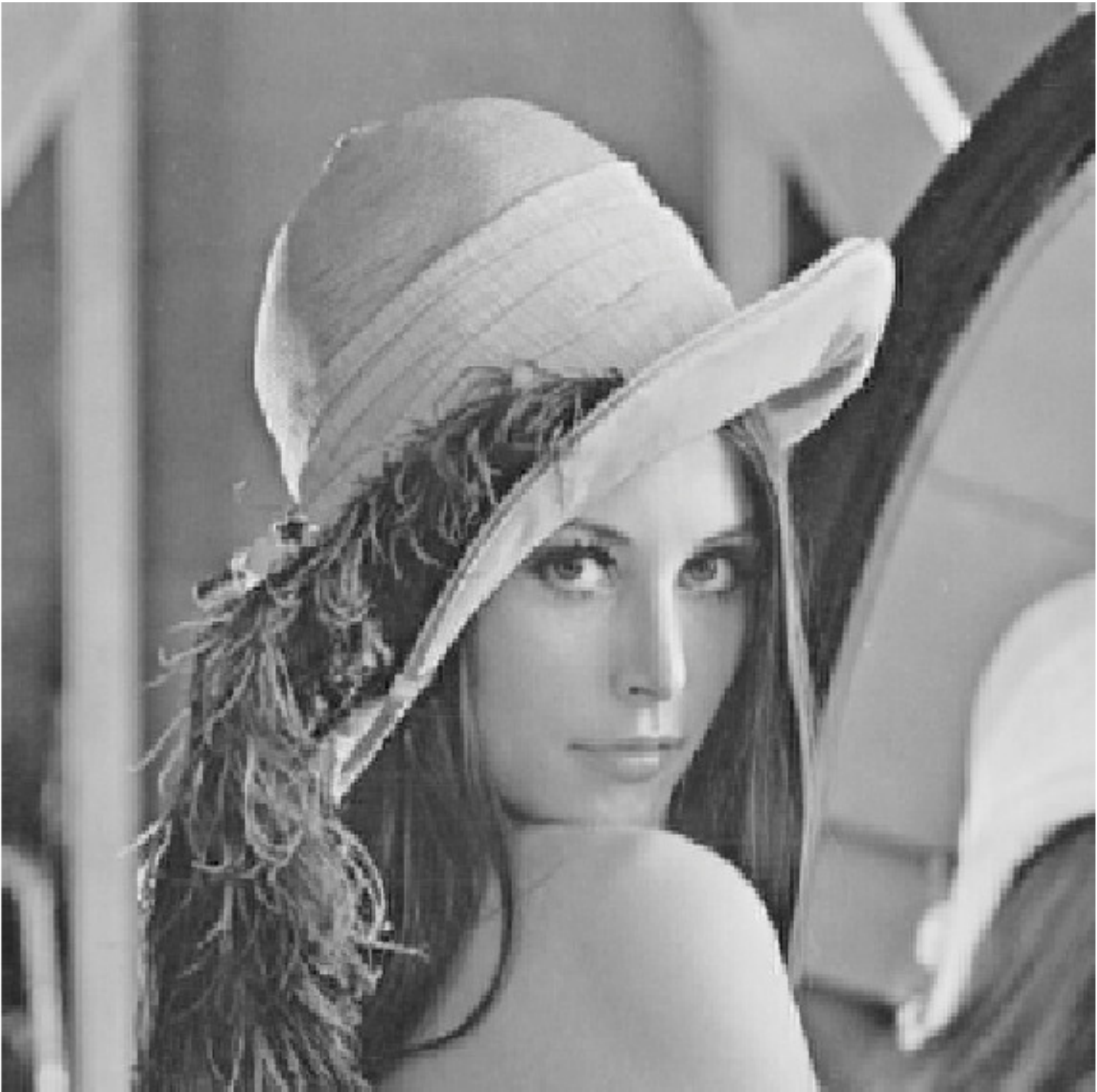} \label{lena_t2}}
\subfigure[$\bm{\mathsf{T}}^{(3)}_8$ (PSNR=31.838, SSIM=0.970)]
{\includegraphics[width=0.18\linewidth]{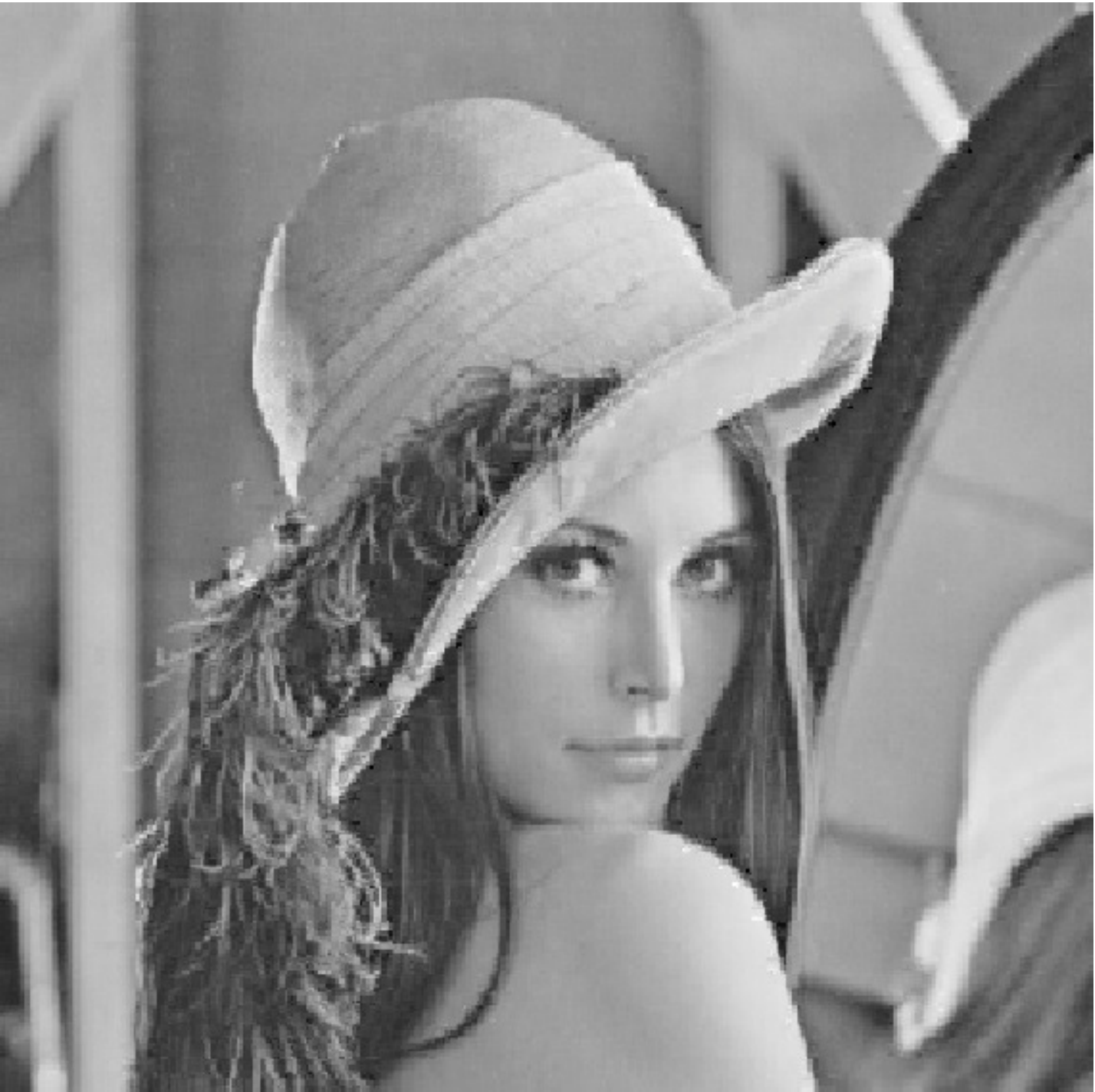} \label{lena_t3}}
\subfigure[$\bm{\mathsf{T}}^{(4)}_8$ (PSNR=32.299, SSIM=0.977)]
{\includegraphics[width=0.18\linewidth]{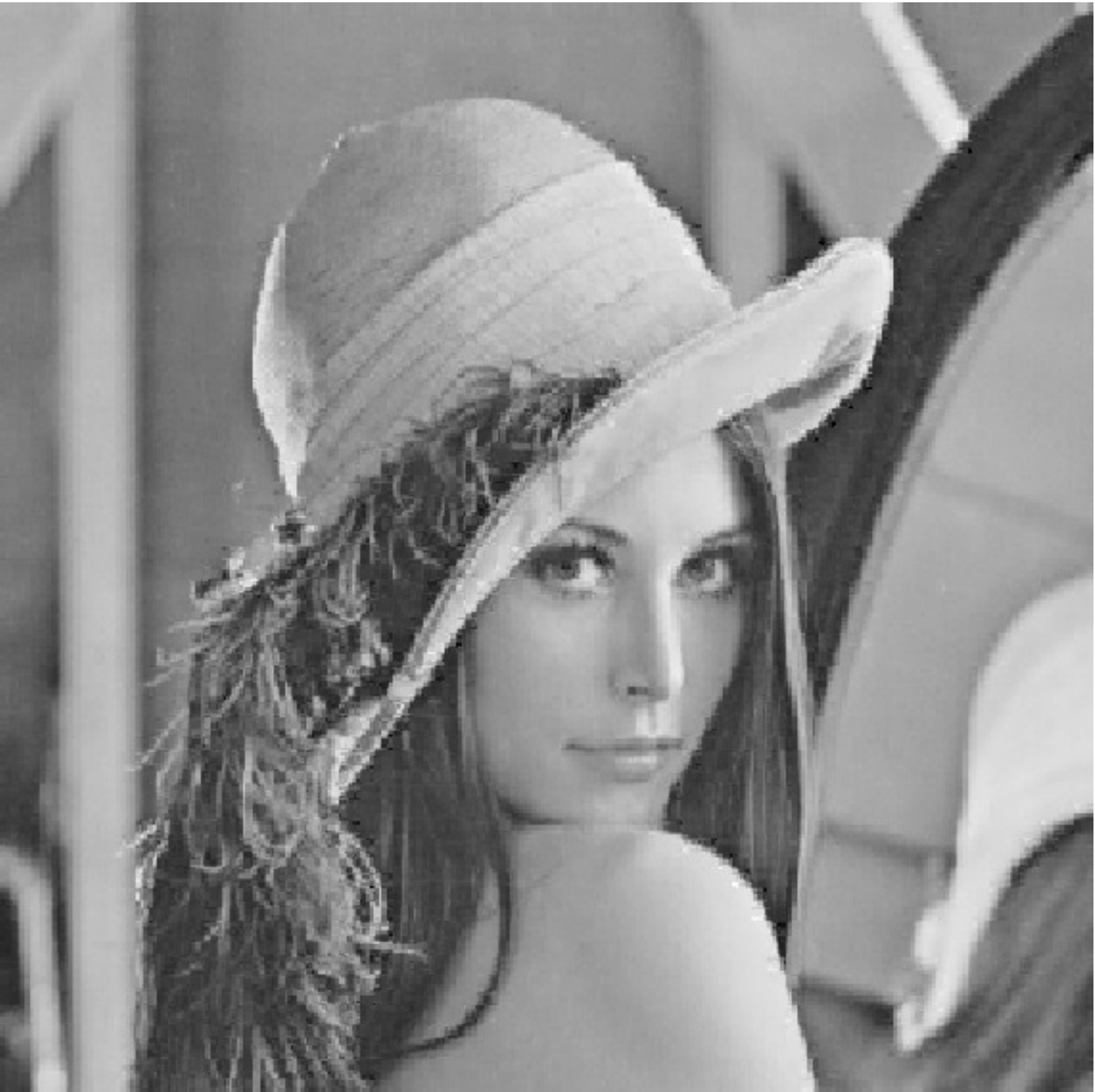} \label{lena_t4}}
\subfigure[$\bm{\mathsf{T}}^{(16)}_8$ (PSNR=31.602, SSIM=0.985)]
{\includegraphics[width=0.18\linewidth]{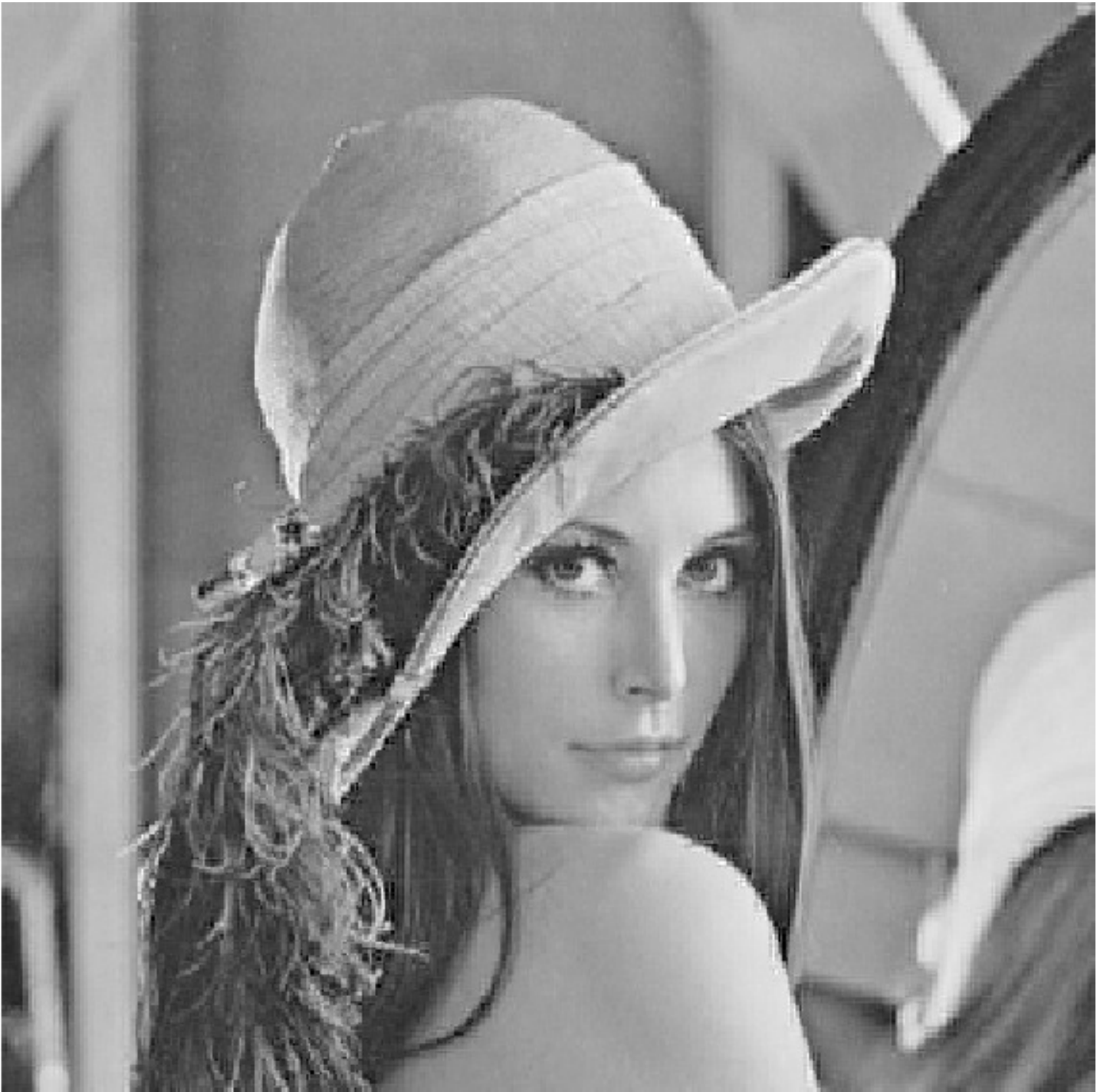} \label{lena_t16}}
\subfigure[$\mbox{BAS}^{(1)}$ (PSNR=31.452, SSIM=0.980)]
{\includegraphics[width=0.18\linewidth]{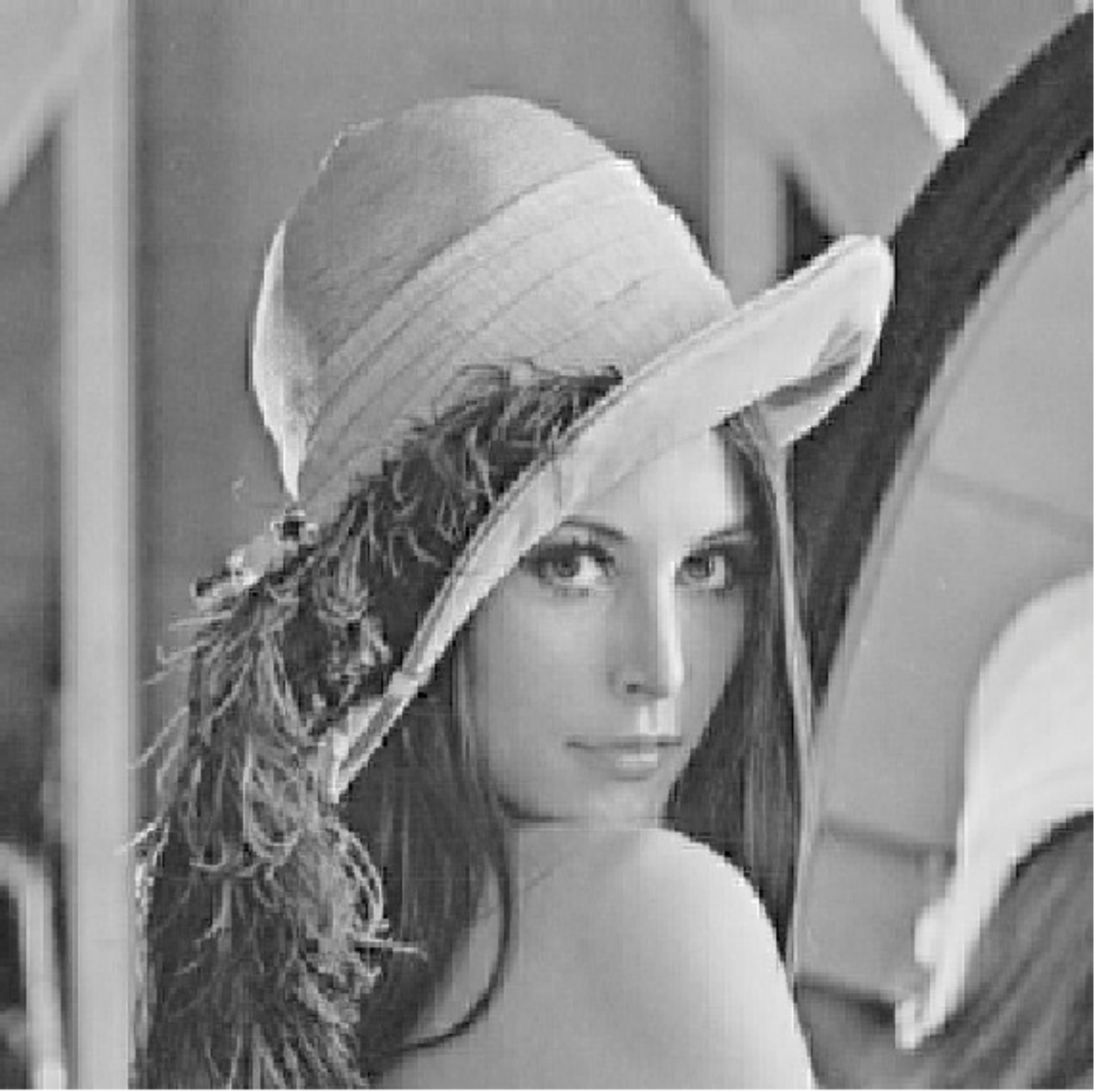} \label{lena_bas1}}
\subfigure[$\mbox{BAS}^{(2)}$ (PSNR=34.794, SSIM=0.991)]
{\includegraphics[width=0.18\linewidth]{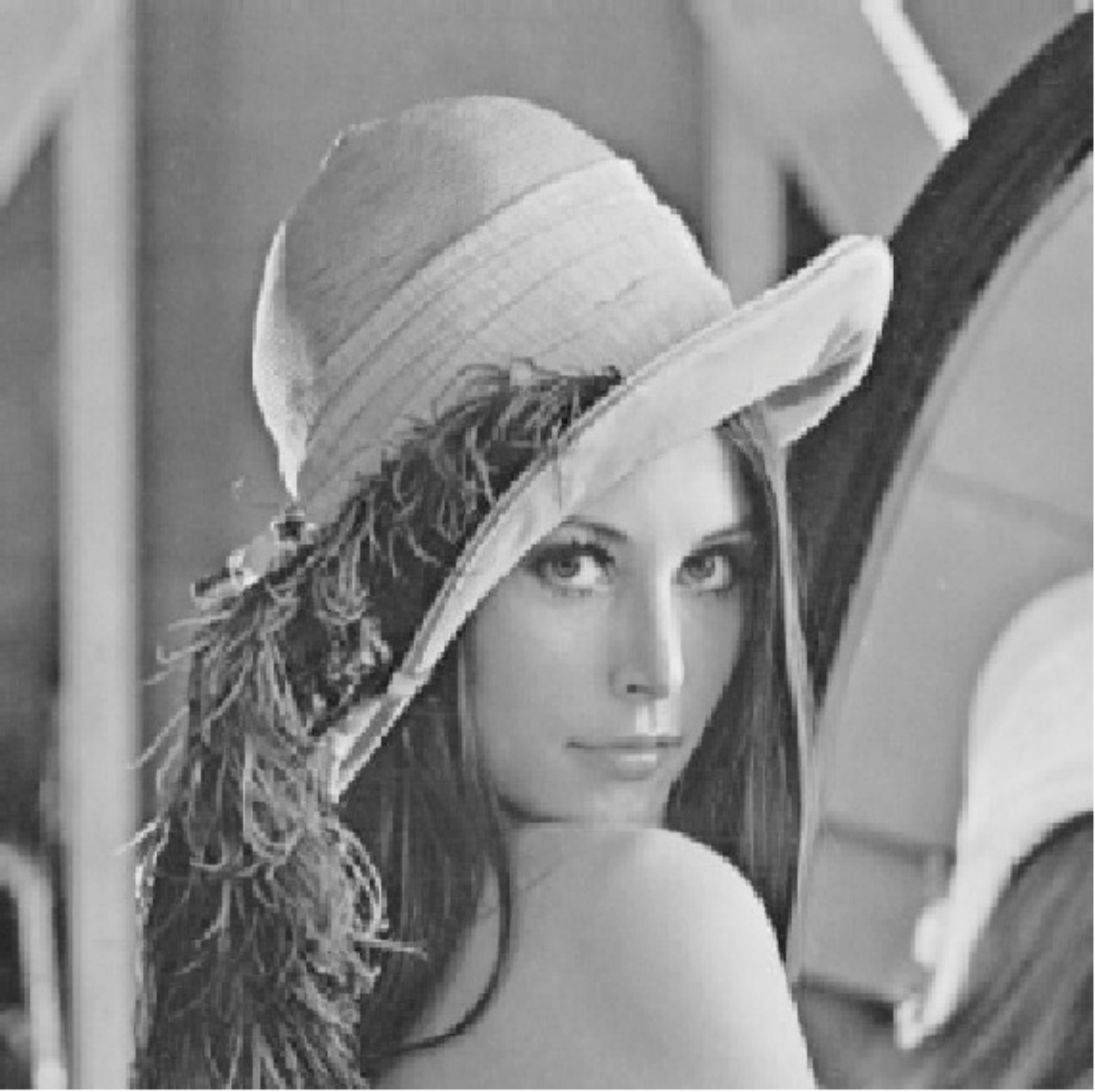} \label{lena_bas2}}
\subfigure[$\mbox{BAS}^{(6)}$ (PSNR=33.819, SSIM=0.986)]
{\includegraphics[width=0.18\linewidth]{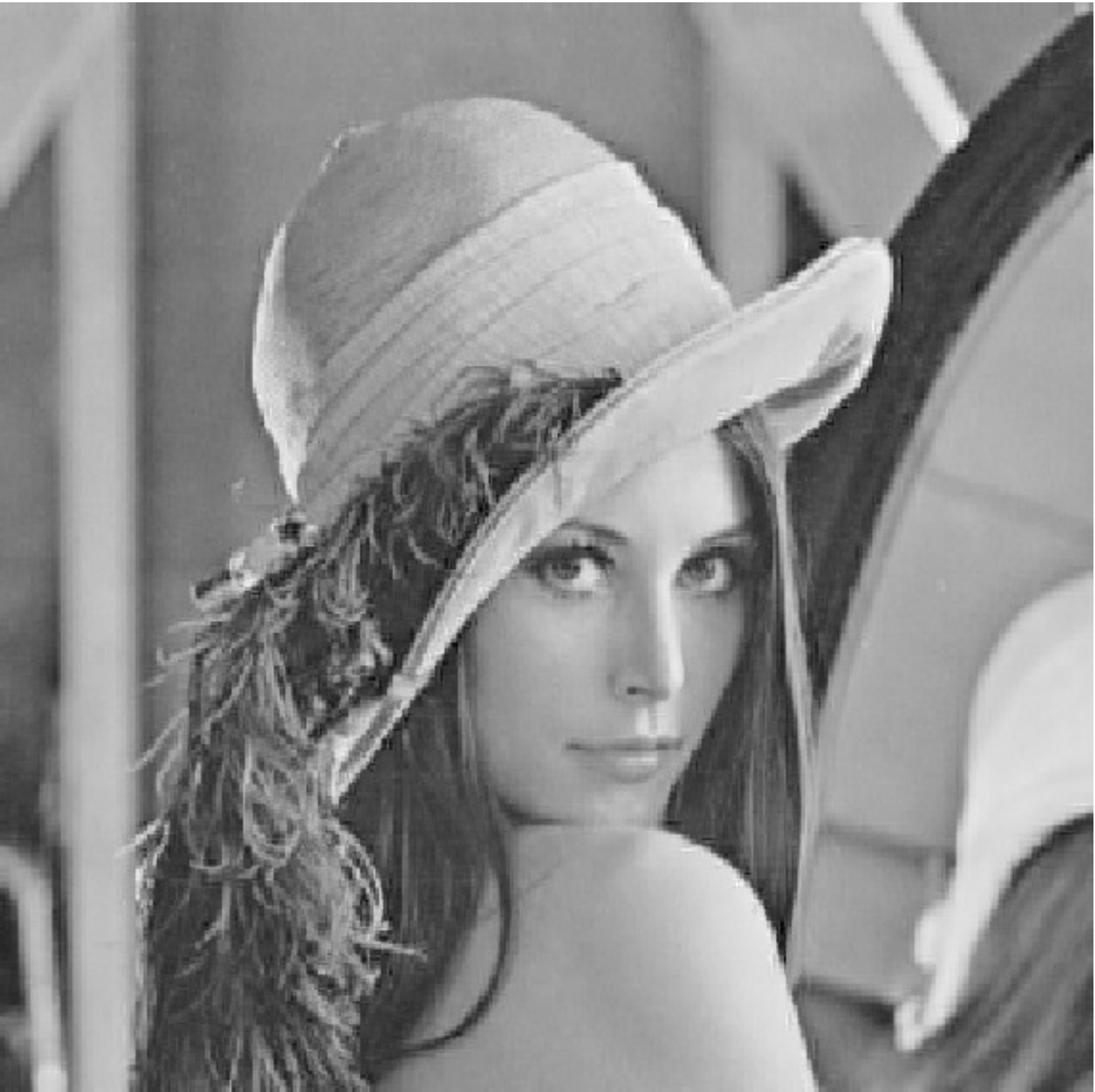} \label{lena_bas6}}
\subfigure[$\mbox{BAS}^{(7)}$ (PSNR=35.433, SSIM=0.995)]
{\includegraphics[width=0.18\linewidth]{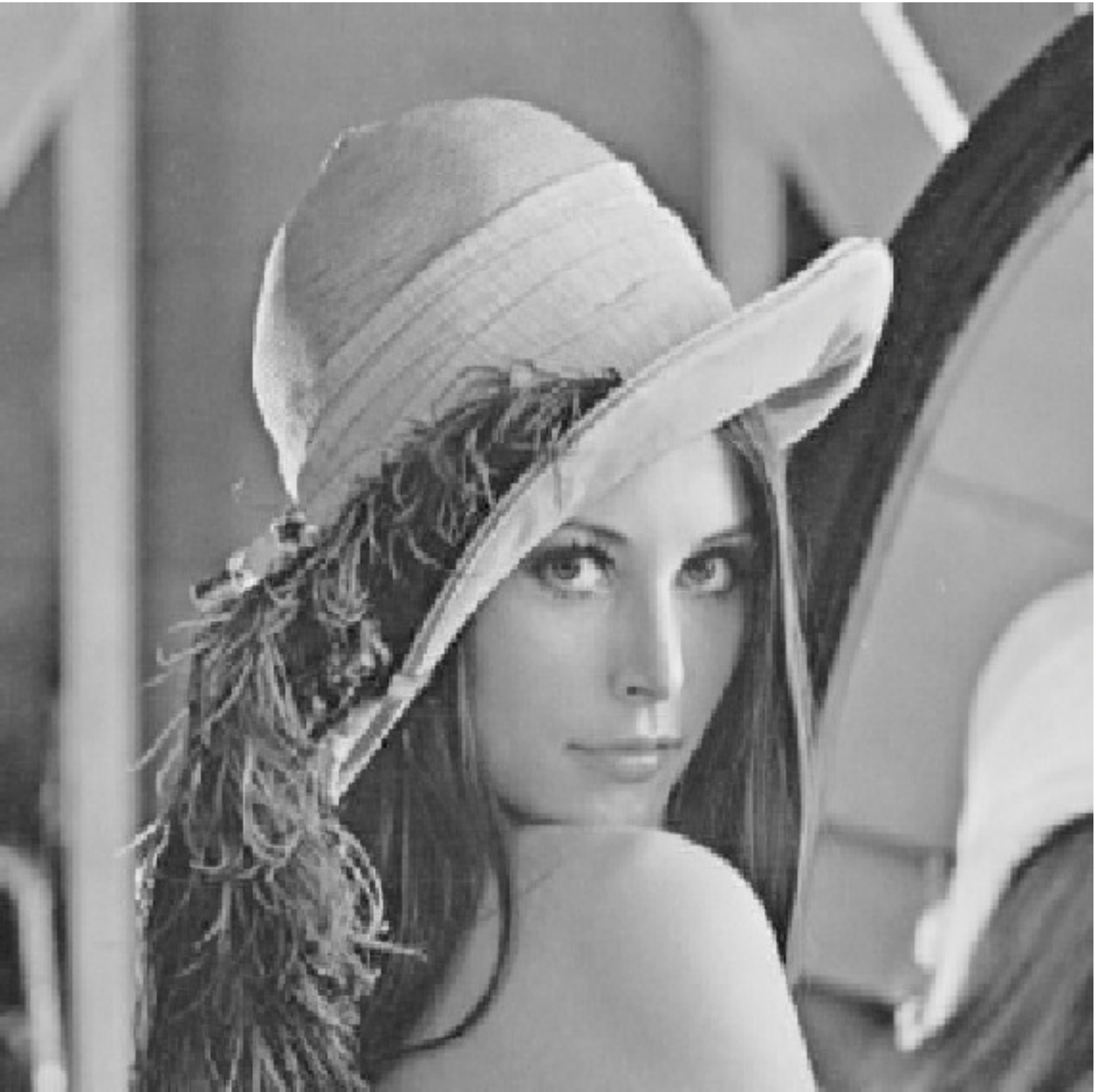} \label{lena_bas7}}
\subfigure[DCT (PSNR=37.886, SSIM=0.997)]
{\includegraphics[width=0.18\linewidth]{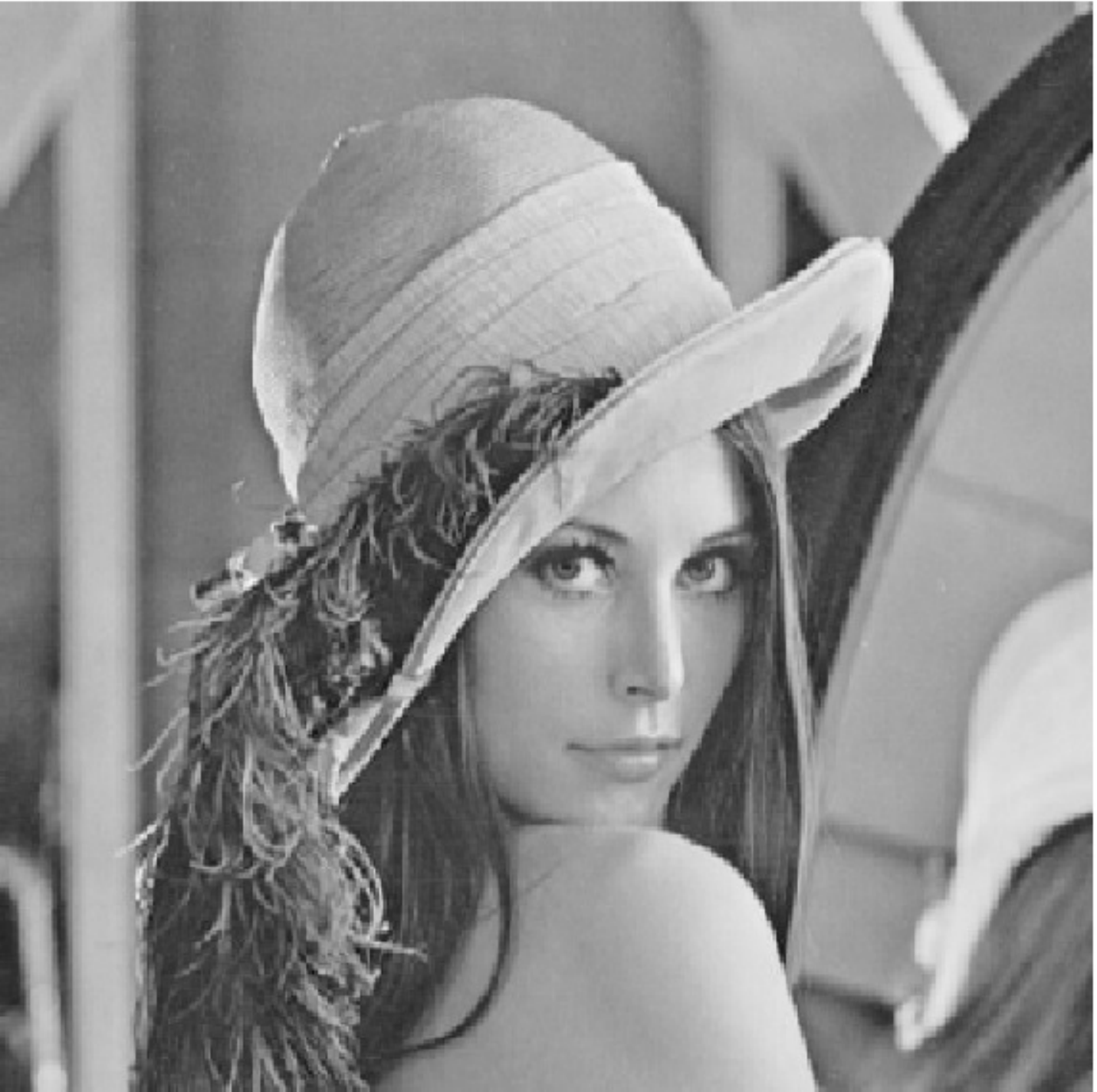} \label{lena_dct}}
\caption{
Compressed Lena image using \subref{lena_t1} $\bm{\mathsf{T}}^{(1)}_8$, \subref{lena_t2} $\bm{\mathsf{T}}^{(2)}_8$,
\subref{lena_t3} $\bm{\mathsf{T}}^{(3)}_8$, \subref{lena_t4} $\bm{\mathsf{T}}^{(4)}_8$, \subref{lena_t16} $\bm{\mathsf{T}}^{(16)}_8$, \subref{lena_bas1} $\mbox{BAS}^{(1)}$, \subref{lena_bas2} $\mbox{BAS}^{(2)}$, \subref{lena_bas6} $\mbox{BAS}^{(6)}$, \subref{lena_bas7} $\mbox{BAS}^{(7)}$, and \subref{lena_dct} DCT, for~$r = 25$~(3.125~bpp).}
\label{f:lena}
\end{figure*}

Fig.~\ref{f:lena} 
shows
512$\times$512
Lena 
image
after 
being submitted to
the described JPEG-like compression,
which provides
a qualitative comparison 
for $r=25$ (3.125~bpp).
This corresponds
to
discarding $60\%$ 
of the 
approximate DCT coefficients. 
PSNR and SSIM measurements are included for comparison.

\section{Conclusion and final remarks}
\label{S:conclusion}

We introduced a new class of matrices based on
a parametrization of the
Feig-Winograd DCT factorization.
By solving a constrained multicriteria optimization problem,
several low-complexity DCT approximations were obtained.
Among the obtained solutions,
we identified DCT approximations already reported in literature.
Thus our procedure furnishes a mathematical framework
that mathematically unifies them.
Moreover,
we derived novel DCT approximations.
The new transformations were 
assessed in terms of
computational complexity,
DCT matrix proximity,
coding gain,
and 
performance in 
JPEG-like compression.
The DCT approximations in the Feig-Winograd matrix subspace
exhibited roughly a trade-off between
cost and performance.
The computational complexity of the proposed transforms
ranged 
from 14~to 24~additions
and
from 0~to 2~bit-shifting operations. 
It is worth to note that
all transforms in the Feig-Winograd class
possess the same algorithm structure. 
Furthermore,
the associated inverse transforms
share 
similar mathematical formalism
and possess simple fast algorithms.
Thus,
in terms of circuitry design,
one can interchange transforms with minimal 
hardware modifications.
In emerging reconfigurable systems, 
it may be possible
to switch \emph{modus operandi} based on the demanded picture
quality and required energy consumption.
Thus, 
the proposed class of approximations
may be a candidate suite of fast algorithms
in such context.
Besides the image compression context,
the Feig-Winograd class of DCT approximations 
can be applied for data encryption
following the scheme introduced in~\cite{Chen2006,bas2010encryption}. 
Finally,
we remark that the proposed method is sufficiently flexible
to be extended to large blocklengths that are powers of two,
according to the Feig-Winograd general theory for DCT factorizaton~\cite[p.~2188]{fw1992}.

\section*{Acknowledgments}

This research was partially supported by CNPq, CAPES, FACEPE, and FAPERGS, Brazil.

{\small
\bibliographystyle{IEEEtran}
\bibliography{fw_ref}
}

\end{document}